\documentclass[11pt]{article}
\pdfoutput=1
\usepackage{jheppub}
\usepackage{graphicx}
\usepackage{amssymb}
\usepackage{amsmath}
\usepackage{amsfonts}
\usepackage{latexsym}
\usepackage{hyperref}
\usepackage{physics}
\usepackage{bbm}
\usepackage{empheq}
\usepackage{tikz}
\usepackage{cancel}
\usepackage{hhline}
\usepackage{caption}
\usepackage{soul}

\usetikzlibrary{arrows.meta}
\usetikzlibrary{decorations.markings}
\tikzset{->-/.style={decoration={
			markings,
			mark=at position #1 with {\arrow{>}}},postaction={decorate}}}

\newcommand\nn{\nonumber}
\newcommand\fft[2]{\frac{#1}{#2}}

\newcommand\tu{\tilde{u}}

\newcommand\mW{\mathcal{W}}

\newcommand\mF{\mathcal{F}}

\newcommand\mfa{\mathfrak{a}}

\newcommand\mA{\mathcal{A}}
\newcommand\tzeta{\widetilde{\zeta}}

\newcommand\mn{\mathfrak{n}}

\newcommand\mm{\mathfrak{m}}

\newcommand\mV{\mathcal{V}}

\newcommand\mN{\mathcal{N}}
\def\ri{{\rm i}}
\newcommand\mR{\mathcal{R}}

\newcommand\mO{\mathcal{O}}

\newcommand\mg{\mathfrak{g}}
\newcommand\tPsi{\widetilde{\Psi}}

\newcommand\mM{\mathcal{M}}

\newcommand\mH{\mathcal{H}}
\newcommand\mS{\mathcal{S}}
\newcommand\bu{\boldsymbol{u}}

\newcommand\Li{\text{Li}}

\newcommand\bDelta{\boldsymbol{\Delta}}
\newcommand\bx{\boldsymbol{x}}
\newcommand\by{\boldsymbol{y}}

\newcommand\bmm{\boldsymbol{\mathfrak{m}}}
\newcommand\bmn{\boldsymbol{\mathfrak{n}}}

\newcommand\vol{\text{vol}}
\newcommand\tchi{\widetilde{\chi}}
\newcommand\hA{\hat{A}}
\newcommand\hsigma{\hat{\sigma}}
\newcommand\hD{\hat{D}}
\newcommand\hmA{\hat{\mA}}
\newcommand\bnu{\boldsymbol{\nu}}
\newcommand\mfc{\mathfrak{c}}
\newcommand\hN{\hat{N}}
\newcommand\hf{\hat{f}}
\newcommand\hg{\hat{g}}

\makeatletter
\newcommand*{\rom}[1]{\expandafter\@slowromancap\romannumeral #1@}
\makeatother

\begin{document}

\title{Perturbatively exact supersymmetric partition functions of ABJM theory on Seifert manifolds and holography}
	
\author[a]{Junho Hong}

\affiliation[a]{Department of Physics \& Center for Quantum Spacetime, Sogang University\,,\\ 35 Baekbeom-ro, Mapo-gu, Seoul 04107, Republic of Korea}

\emailAdd{junhohong@sogang.ac.kr}

	
\abstract{We undertake a comprehensive analysis of the supersymmetric partition function of the $\text{U}(N)_k\times\text{U}(N)_{-k}$ ABJM theory on a U(1) fibration over a Riemann surface, evaluating it to all orders in the $1/N$-perturbative expansion up to exponentially suppressed corrections. Through holographic duality, our perturbatively exact result is successfully matched with the regularized on-shell action of a dual Euclidean AdS$_4$-Taub-Bolt background incorporating 4-derivative corrections, and also provides valuable insights into the logarithmic corrections that emerge from the 1-loop calculations in M-theory path integrals. In this process, we revisit the Euclidean AdS$_4$-Taub-Bolt background carefully, elucidating the flat connection in the background graviphoton field. This analysis umambiguously determines the U(1)$_R$ holonomy along the Seifert fiber, thereby solidifying the holographic comparison regarding the partition function on a large subclass of Seifert manifolds.}
	
\maketitle \flushbottom

\section{Introduction}\label{sec:intro}
Supersymmetric localization \cite{Pestun:2007rz,Pestun:2016zxk} has emerged as an exceptionally powerful technique, enabling our community to investigate partition functions and integrated correlators of supersymmetric field theories through the corresponding matrix models. This approach becomes particularly illuminating when applied to holographic theories, as their partition functions and integrated correlators serve as invaluable tools for examining holographic dual path integrals in string/M-theory.

The $S^3$ partition function of the $\text{U}(N)_k\times\text{U}(N)_{-k}$ ABJM theory \cite{Aharony:2008ug} has been the subject of extensive investigation within this framework. Based on the matrix integral formulation obtained via supersymmetric localization \cite{Kapustin:2009kz}, the ABJM $S^3$ partition function was evaluated in the large $N$ limit, where the characteristic $N^{3/2}$ leading order \cite{Klebanov:1996un} was successfully matched with the regularized on-shell action of a dual Euclidean AdS$_4$ background through the AdS/CFT correspondence \cite{Drukker:2010nc,Herzog:2010hf}. Further developments revealed that the matrix model for the ABJM $S^3$ partition function could be computed to all orders in the $1/N$-perturbative expansion through a series of dualities involving the topological string on the canonical line bundle over $\mathbb{P}^1\times \mathbb{P}^1$ \cite{Fuji:2011km} or through the free Fermi-gas formulation \cite{Marino:2011eh}. The result is elegantly encapsulated by an Airy function, whose $1/N$-perturbative expansion has significantly enlightened our comprehension of a dual Euclidean M-theory path integral beyond the semi-classical limit, where the approximation by the regularized on-shell action of a dual Euclidean AdS$_4$ background requires refinement. Specifically, the $N^{1/2}$ order corrections have been instrumental in exploring higher-derivative corrections to the regularized on-shell action \cite{Bobev:2020egg,Bobev:2021oku}, while the $\log N$ order corrections were successfully reproduced via an 11d supergravity 1-loop calculation \cite{Bhattacharyya:2012ye}.

Recent insights have illuminated that such comprehensive information regarding a 3d supersymmetric partition function of ABJM theory extends beyond the $S^3$ partition function. A pivotal observation is that the $S^1\times\Sigma_{\mg}$ partition function of ABJM theory with a partial topological twist over the Riemann surface $\mg$, referred to as the topologically twisted index (TTI) \cite{Benini:2015noa,Benini:2015eyy}, can be expressed in a succinct closed-form expression valid to all orders in the $1/N$-perturbative expansion \cite{Bobev:2022jte,Bobev:2022eus}. This all-order TTI serves as a foundation for exploring the superconformal index (SCI) of the same theory, wherein the leading two terms of the Cardy-like expansion are similarly derived to all orders in the $1/N$-perturbative expansion \cite{Bobev:2022wem}. These findings extend to a broader class of $\mN\geq2$ holographic superconformal field theories (SCFTs) arising from $N$ coincident M2 branes probing the orbifold singularities of Calabi-Yau 4-folds as well as the real-mass deformed ABJM theory \cite{Bobev:2023lkx,Bobev:2024mqw}. Moreover, previous analyses of their $S^3$ partition functions have been refined in this line of research by extending the Airy formula beyond its original limitations \cite{Fuji:2011km,Marino:2011eh,Mezei:2013gqa,Grassi:2014vwa,Nosaka:2015iiw,Hatsuda:2021oxa,Chester:2023qwo} to encompass a wider range of cases \cite{Bobev:2022jte,Bobev:2022eus,Hristov:2022lcw,Bobev:2023lkx,Geukens:2024zmt,Hong:2024}. Consequently, this breakthrough paves the way for a comprehensive exploration of perturbatively exact 3d supersymmetric partition functions in $\mN\geq2$ holographic SCFTs dual to M-theory, along with the associated holographic analyses.

In this paper we further advance the aforementioned accomplishment by analyzing the supersymmetric partition function of ABJM theory on a large class of Seifert manifolds to all orders in the $1/N$-perturbative expansion and investigating its holographic implications. In section \ref{sec:Mgp}, we review the 3d supersymmetric background corresponding to a U(1) bundle over a Riemann surface of genus $\mg$ with first Chern number $p$, representing a class of Seifert manifolds denoted by $\mM_{\mg,p}$, and summarize the Bethe formulation for the partition function of $\mN=2$ Chern-Simons-matter theories on this supersymmetric background \cite{Closset:2017zgf,Closset:2018ghr,Closset:2019hyt}. In section \ref{sec:ABJM}, we apply the Bethe formulation to the $\text{U}(N)_k\times\text{U}(N)_{-k}$ ABJM theory, evaluating its $\mM_{\mg,p}$ partition function to all orders in the $1/N$-perturbative expansion, which significantly enhances the prior work limited to the $N^{3/2}$ leading order \cite{Toldo:2017qsh}. This is achieved by employing the closed-form expressions for the Bethe potential and the TTI of the same theory, as determined through numerical analysis in \cite{Bobev:2022jte,Bobev:2022eus,Bobev:2022wem}. Section \ref{sec:sugra} revisits the Euclidean AdS-Taub-Bolt geometry that serves as the holographic dual to the $\mM_{\mg,p}$ partition function of interest \cite{Toldo:2017qsh,Bobev:2020zov,Martelli:2012sz,Chamblin:1998pz,Hawking:1998ct,Alonso-Alberca:2000zeh,Taub:1950ez,Newman:1963yy}, clarifying the flat connection associated with the background graviphoton inspired by insights from \cite{Cabo-Bizet:2018ehj,BenettiGenolini:2023rkq,BenettiGenolini:2023ucp}. In section \ref{sec:holo}, we first identify the $\mM_{\mg,p}$ background and the asymptotic behavior of the dual Euclidean AdS-Taub-Bolt geometry, while specifying the U(1)$_R$ holonomy in the boundary field theory as informed by the graviphoton flat connection. We then proceed to compare the $\mM_{\mg,p}$ partition function of ABJM theory with the dual Euclidean M-theory path integral beyond the strict large $N$ limit, offering highly nontrivial support for the supergravity analysis dual to the $N^{1/2}$ \& $\log N$ order corrections \cite{Bobev:2020egg,Bobev:2021oku,Bobev:2020zov,Hristov:2021zai,Bobev:2023dwx}. This holographic comparison is executed focusing on the universal twist \cite{Azzurli:2017kxo,Bobev:2017uzs}, which is unequivocally determined by the specific values of U(1)$_R$ holonomy. This aspect has been overlooked in the literature concerning the holography of the $\mM_{\mg,p}$ partition function with a non-vanishing $p$. In section \ref{sec:discussion}, we summarize our findings and discuss potential future directions.

\section{Supersymmetric partition functions on Seifert manifolds}\label{sec:Mgp}
This section reviews the supersymmetric background on a large class of Seifert manifolds, denoted by $\mM_{\mg,p}$, for $\mN=2$ field theories with U(1)$_R$ symmetry and their partition functions. In subsection \ref{sec:Mgp:bg}, we begin by describing the supersymmetric background on $\mM_{\mg,p}$ \cite{Closset:2016arn,Closset:2017zgf,Closset:2018ghr,Closset:2019hyt}, which is constructed using a general formalism that employs the rigid limit of 3d new minimal supergravity \cite{Closset:2012ru,Closset:2013vra}. In subsection \ref{sec:Mgp:ptnfct}, we present the Bethe formulation for the $\mM_{\mg,p}$ partition function of $\mN=2$ field theories with U(1)$_R$ symmetry \cite{Closset:2016arn,Closset:2017zgf,Closset:2018ghr,Closset:2019hyt}.

\subsection{3d supersymmetric background on $\mM_{\mg,p}$}\label{sec:Mgp:bg}
\subsubsection{New minimal supergravity multiplet}\label{sec:Mgp:bg:new-min}
3d $\mN=2$ supersymmetric field theories with U(1)$_R$ symmetry can be placed on Riemannian three-manifolds $\mM_3$ by taking the rigid limit of 3d new minimal supergravity. The 3d new minimal supergravity multiplet comprises two gravitini $\Psi_\mu,\tPsi_\mu$ and bosonic fields
\begin{equation}
	g_{\mu\nu}\,,\quad A_\mu^{(R)},\,\quad V_\mu,\,\quad H\label{new-minimal}
\end{equation}
that couple to the supercurrent multiplet \cite{Closset:2012ru}, where we have followed the conventions of \cite{Closset:2017zgf,Closset:2018ghr,Closset:2019hyt}. In the process of taking the rigid limit, a portion of the original $\mN=2$ supersymmetry can be preserved by appropriately choosing the background values of the bosonic fields (\ref{new-minimal}) to solve the generalized Killing spinor equations
\begin{subequations}
	\begin{align}
		0&=\bigg(\partial_\mu+\fft14\omega_\mu^{ab}\gamma_{ab}-\ri A_\mu^{(R)}+\fft12H\gamma_\mu-\fft{\ri}{2}V_\mu+\fft12\epsilon_{\mu\nu\rho}V^\nu\gamma^\rho\bigg)\zeta\,,\\
		0&=\bigg(\partial_\mu+\fft14\omega_\mu^{ab}\gamma_{ab}+\ri A_\mu^{(R)}+\fft12H\gamma_\mu+\fft{\ri}{2}V_\mu-\fft12\epsilon_{\mu\nu\rho}V^\nu\gamma^\rho\bigg)\tzeta\,.
	\end{align}\label{3d:KSE}%
\end{subequations}
In this paper, we focus specifically on the case of a Seifert manifold $\mM_3=\mM_{\mg,p}$, which is a U(1) principal bundle over the Riemann surface $\Sigma_{\mg}$ of genus $\mg$ with first Chern number $p$, where two real supercharges are present. Below, we briefly review this 3d supersymmetric background \cite{Closset:2017zgf,Toldo:2017qsh} using slightly different conventions.

\medskip

The 3d metric on the Seifert manifold can be expressed as
\begin{equation}
	ds_3^2=\fft{1}{4\eta} ds^2_{\Sigma_{\mg}}+\beta_\text{3d}^2(d\psi+p\mfa)^2\,,\label{Seifert:metric}
\end{equation}
where the parameter $\eta$, the 1-form $\mfa$, and the Riemann surface metric $ds^2_{\Sigma_{\mg}}$ are defined as (with $\kappa$ taking values $\kappa\in\{1,0,-1\}$ for $\mg\in\{0,1,{>}1\}$ respectively)
\begin{equation}
\begin{alignedat}{2}
	\eta&=\begin{cases}
		|\mg-1| & (\mg\neq1)\\
		1 & (\mg=1)
	\end{cases}\,,&\qquad f_\kappa(\theta)&=\begin{cases}
	-\cos\theta & (\kappa=1) \\
	\theta & (\kappa=0) \\
	\cosh\theta & (\kappa=-1)
	\end{cases}\,,\\
	\mfa&=\fft{1}{2\eta}f_\kappa(\theta)d\phi\,,&\qquad
	ds^2_{\Sigma_{\mg}}&=d\theta^2+f_\kappa'(\theta)^2d\phi^2\,.
\end{alignedat}\label{f:Sigma}%
\end{equation}
The volume of the Riemann surface $ds^2_{\Sigma_\mg}$ is given by $\vol[\Sigma_\mg]=4\pi\eta$; the prefactor $\fft{1}{4\eta}$ normalizes the area of the 2-dimensional surface in (\ref{Seifert:metric}) to $\pi$, which is consistent with the conventions of \cite{Closset:2017zgf,Toldo:2017qsh,Closset:2018ghr,Closset:2019hyt}. The fiber coordinate ranges over $\psi\in[0,2\pi)$ and the 1-form $\mfa$ satisfies\footnote{To ensure that the angular coordinate $\phi$ has $2\pi$ periodicity in all cases, we set the range of the $\theta$ coordinate as $\theta\in[0,2)$ for the torus with $\mg=1$. This choice differs from the convention in \cite{Toldo:2017qsh} by a factor of two.}
\begin{equation}
	\fft{1}{2\pi}\int_{\Sigma_{\mg}}d\mfa=1\,.
\end{equation}
As a consequence, the 3d manifold equipped with the metric (\ref{Seifert:metric}) describes a U(1) bundle over a Riemann surface of genus $\mg$ with first Chern number $p\in\mathbb{Z}$. 

\medskip

The background values of the other bosonic fields in the new minimal supergravity multiplet (\ref{new-minimal}) are given by
\begin{subequations}
\begin{align}
	A^{(R)}&=p\beta_\text{3d}^2(d\psi+p\mfa)-\kappa\eta \mfa+d\ell\,,\label{Seifert:AHV:A}\\
	H&=-\ri p\beta_\text{3d}\,,\\
	V_\mu&=0\,,
\end{align}\label{Seifert:AHV}%
\end{subequations}
where we have fixed a free parameter to turn off $V_\mu$, see \cite{Toldo:2017qsh} for further details. In the $R$ symmetry background gauge field (\ref{Seifert:AHV:A}), we have included a flat connection $d\ell=d\ell(\theta,\phi,\psi)$ in the last term that will be crucial below. 

\medskip

The Killing spinor equations on the background (\ref{Seifert:metric}) and (\ref{Seifert:AHV}) allow for solutions
\begin{equation}
	\zeta=e^{\ri \ell}\begin{pmatrix}
		0 \\ \chi
		\end{pmatrix}\quad\&\quad \tzeta=e^{-\ri \ell}\begin{pmatrix}
		\tchi \\ 0
		\end{pmatrix}\qquad(\chi,\tchi\in\mathbb{C})\,,\label{KS}
\end{equation}
where we have employed the local orthonormal frame
\begin{equation}
	e^1=\fft{1}{2\sqrt{\eta}}d\theta\,,\qquad e^2=\fft{1}{2\sqrt{\eta}}f_\kappa'(\theta)d\phi\,,\qquad e^3=\beta_\text{3d}(d\psi+p\mfa)\,,\label{3d:lof}
\end{equation}
for the metric (\ref{Seifert:metric}). Identifying the nowhere vanishing Killing vector $K^\mu=\zeta\gamma^\mu\tzeta$ made out of the Killing spinors (\ref{KS}) with the U(1) isometry in the background (\ref{Seifert:metric}) as $K=\partial_\psi$, we find that the complex parameters $\chi,\tchi$ are related by
\begin{equation}
	\tchi=\beta_\text{3d}/\chi\,.\label{chi:tchi}
\end{equation}
Consequently, the Seifert manifold preserves two real supercharges \cite{Closset:2012ru}. For conventions used in deriving the Killing spinors (\ref{KS}) and the relation (\ref{chi:tchi}), see Appendix \ref{app:conventions}.

\subsubsection{Background vector multiplets}\label{sec:Mgp:bg:vec}
Next we introduce various parameters used in the next subsection \ref{sec:Mgp:ptnfct} summarizing the Bethe formulation. To this end, we first consider a sub-multiplet of the full new minimal supergravity multiplet that corresponds to the abelian background vector multiplet associated with the U(1)$_R$ symmetry. The bosonic content of this sub-multiplet is given in terms of the supergravity bosonic fields (\ref{new-minimal}) as \cite{Closset:2012ru,Closset:2019hyt} 
\begin{equation}
	\hA=A^{(R)}+V\,,\qquad \hsigma=H\,,\qquad \hD=\fft14(R+2H^2+2V_\mu V^\mu)\,.
\end{equation}
We then introduce the complexified $R$ symmetry background gauge field as \cite{Closset:2018ghr} 
\begin{equation}
	\hmA=\hA-\ri\hsigma e^3\label{hmA:R}
\end{equation}
in terms of the local orthonormal frame (\ref{3d:lof}). The magnetic flux associated with the $R$ symmetry is defined as
\begin{equation}
	\mn_R\equiv\fft{1}{2\pi}\int_{\Sigma_\mg}d\hmA\,.\label{n:R}
\end{equation}
The U(1)$_R$ holonomy of the complexified $R$ symmetry background gauge field (\ref{hmA:R}) along the Seifert fiber, denoted by $\gamma$, defines the complex modulus as \cite{Closset:2018ghr}
\begin{equation}
	\nu_R\equiv-\fft{1}{2\pi}\int_{\gamma}\hmA\,.\label{nu:R}
\end{equation}
Throughout this paper we assume that the $\ell$-function governing the flat connection takes the form $\ell=-\nu_R\psi$, a choice that is consistent with the U(1)$_R$ holonomy (\ref{nu:R}) and determines the magnetic flux (\ref{n:R}) as $\mn_R=\mg-1$ for the background fields (\ref{Seifert:AHV}). The U(1)$_R$ holonomy $\nu_R$ is supposed to take integer or half-integer values, which is correlated with periodic or anti-periodic boundary conditions for fermions along the Seifert fiber respectively \cite{Closset:2018ghr,Closset:2019hyt}. In section \ref{sec:holo}, we will demonstrate that the dual 4-dimensional supergravity background specifies the U(1)$_R$ holonomy as $\nu_R\in\{\pm\fft12\}$, thereby requiring anti-periodic fermions.

\medskip

If a given $\mN=2$ field theory has any other U(1) flavor symmetry, we may also consider the associated abelian vector multiplet with the bosonic content
\begin{equation}
	\hA^{(F)}\,,\qquad \hsigma^{(F)}\,,\qquad \hD^{(F)}\,.\nn
\end{equation}
The complexified flavor symmetry background gauge field is then introduced as
\begin{equation}
	\hmA^{(F)}=\hA^{(F)}-\ri\hsigma^{(F)} e^3\,,\label{hmA:F}
\end{equation}
with the corresponding magnetic flux and the holonomy defined as in (\ref{n:R}) and (\ref{nu:R}) as
\begin{equation}
	\mn_F\equiv\fft{1}{2\pi}\int_{\Sigma_{\mg}}d\hmA^{(F)}\,,\qquad \nu_F\equiv-\fft{1}{2\pi}\int_\gamma\hmA^{(F)}\,,\label{n:nu:F}
\end{equation}
respectively.

\subsection{3d supersymmetric partition function on $\mM_{\mg,p}$: Bethe formulation}\label{sec:Mgp:ptnfct}
The partition function of a 3d $\mN=2$ field theory with U(1)$_R$ symmetry on the supersymmetric background $\mM_{\mg,p}$ described in the previous subsection \ref{sec:Mgp:bg}, has been investigated through the Kaluza-Klein (KK) compactification along the Seifert fiber that yields a 2d $\mN=(2,2)$ theory with infinitely many fields \cite{Closset:2016arn,Closset:2017zgf,Closset:2018ghr,Closset:2019hyt}. In this framework, the partition function can be written as a sum over Bethe vacua with the building blocks governed by the effective twisted superpotential and the effective dilaton. We refer the readers to the literature \cite{Closset:2016arn,Closset:2017zgf,Closset:2018ghr,Closset:2019hyt} for further details. Below we simply summarize the resulting Bethe formulation for the 3d supersymmetric partition function on $\mM_{\mg,p}$.

\medskip

Before presenting the Bethe formulation for the $\mM_{\mg,p}$ partition function, we first outline the conventions describing $\mN=2$ Chern-Simons-matter theories with the gauge group $G$ and the flavor group $H$, where the matter content is collectively represented by $\mN=2$ chiral multiplets $\Psi$ with the $R$-charge $r_\Psi$.
\begin{itemize}
	\item $i,j\in\{1,2,\cdots,|G|\}$ and $I,J\in\{1,2,\cdots,|H|\}$ where $|G|$ and $|H|$ represent the rank of the gauge/flavor groups respectively. $W$ denotes the Weyl transformation group associated with the gauge group $G$, whose rank is represented by $|W|$.
	
	\item $x_i=e^{2\pi\ri u_i}$ and $y_I=e^{2\pi\ri \nu_I}$ denote the exponentiation of gauge/flavor holonomies respectively, see (\ref{n:nu:F}) for the introduction of the $\nu_I$ parameter for example. $\mm_i$ and $\mn_I$ denote gauge/flavor magnetic fluxes respectively. The bold symbols $\bx$ \& $\by$ and $\bmm$ \& $\bmn$ are used to represent them collectively.
	
	\item $k^{ij}$, $k^{iI}$, $k^{iR}$, $k^{IJ}$, $k^{IR}$, and $k^{RR}$ denote the (mixed) Chern-Simons (CS) levels for gauge-gauge, gauge-flavor, gauge-$R$, flavor-flavor, flavor-$R$, and $R$ symmetries respectively. $k^g$ stands for the gravitational CS level.
	
	\item $\rho_\Psi$ \& $\gamma_\Psi$ denote the weights of the representation of an $\mN=2$ chiral multiplet $\Psi$ with respect to the gauge group $G$ \& the flavor group $H$ respectively. They can be understood as linear maps acting on the Cartan subalgebras $\mathfrak{h}_G$ and $\mathfrak{h}_H$ associated with the gauge group $G$ and the flavor group $H$ respectively as 
	\begin{equation}
		\begin{split}
			\rho_\Psi(X)&=\sum_{i=1}^{|G|}(\rho_\Psi)_iX_i \qquad(X=X_i\mH^G_i\in\mathfrak{h}_G)\,,\\
			\gamma_\Psi(Y)&=\sum_{I=1}^{|H|}(\gamma_\Psi)_IY_I \qquad(Y=Y_I\mH^{H}_I\in\mathfrak{h}_H)\,,
		\end{split}
	\end{equation}
	where $\mH^G_i$ and $\mH^{H}_I$ are the corresponding Cartan generators. For notational convenience we use compact expressions
	\begin{equation}
		\bx^{\rho_\Psi}=e^{2\pi\ri \rho_\Psi(\bu)}=e^{2\pi\ri\sum_{i=1}^{|G|}(\rho_\Psi)_iu_i} \quad\&\quad \by^{\gamma_\Psi}=e^{2\pi\ri \gamma_\Psi(\bnu)}=e^{2\pi\ri\sum_{I=1}^{|H|}(\gamma_\Psi)_I\nu_I}\,.
	\end{equation}

	\item $\mR[G]\subseteq(\mathfrak{h}_G)^*$ denotes the roots associated with the gauge group $G$. We use $\alpha\in\mR[G]$ to represent its element, and employ the compact expression
	\begin{equation}
		\bx^\alpha=e^{2\pi\ri\alpha(\bu)}=e^{2\pi\ri\sum_{i=1}^{|G|}(\alpha)_iu_i}\,.
	\end{equation}
\end{itemize}

\medskip

Based on the conventions introduced above and the Bethe formulation from \cite{Closset:2017zgf,Closset:2018ghr,Closset:2019hyt}, the $\mM_{\mg,p}$ partition function of an $\mN=2$ CS-matter theory can be written as
\begin{equation}
	Z_{\mM_{\mg,p}}(\bnu,\bmn)=\sum_{\bu\in\mS_{BE}}\mF(\bu,\bnu)^p\mH(\bu,\bnu)^{\mg-1}\prod_{I=1}^{|H|}\Pi_I(\bu,\bnu)^{\mn_I}\,,\label{Z}
\end{equation}
which is referred to as the Bethe formula. The building blocks and the Bethe vacua $\mS_{BE}$ that govern the Bethe formula are defined as follows. 
\begin{itemize}
	\item The effective twisted superpotential is given by
	\begin{align}
		\mW(\bu,\bnu)&=\fft12\sum_{i,j=1}^{|G|}k^{ij}u_i(u_j+(1+2\nu_R)\delta_{ij})+\sum_{i=1}^{|G|}k^{\ri R}u_i\nu_R \label{W}\\
		&\quad+\fft12\sum_{I,J=1}^{|H|}k^{IJ}\nu_I(\nu_J+(1+2\nu_R)\delta_{IJ})+\sum_{I=1}^{|H|}k^{IR}\nu_I\nu_R+\sum_{i=1}^{|G|}\sum_{I=1}^{|H|}k^{iI}u_i\nu_I+\fft{1}{24}k^g\nn\\
		&\quad+\fft{1}{(2\pi\ri)^2}\sum_\Psi\sum_{\rho_\Psi}\Li_2(\bx^{\rho_\Psi}\by^{\gamma_\Psi}e^{2\pi\ri  r_\Psi\nu_R})+\fft14\sum_\Psi\sum_{\rho_\Psi}B_2([\rho_\Psi(\bu)+\gamma_\Psi(\bnu)+r_\Psi\nu_R])\,,\nn
	\end{align}
	where we have introduced a notation
	\begin{equation}
		[x]\equiv x+1-\lceil\Re[x]\rceil\,.\nn
	\end{equation}
	The last Bernoulli polynomial term in (\ref{W}) is absent in the gauge-invariant quantization \cite{Closset:2017zgf,Toldo:2017qsh,Closset:2018ghr,Closset:2019hyt}, but it is included in the parity-preserving quantization \cite{Benini:2015noa,Hosseini:2016tor,Hosseini:2016ume,Toldo:2017qsh,Bobev:2022eus,Bobev:2023lkx,Bobev:2022wem,Bobev:2024mqw}. While this expression is not completely identical to those present in the Bethe potentials in the latter literature, they are equivalent up to the branch-cut ambiguity of the effective twisted superpotential that is constant or linear with respect to the complex gauge holonomies \cite{Closset:2017zgf,Closset:2018ghr,Closset:2019hyt}. 
	
	It is important to note that the effective twisted superpotential (\ref{W}) is formulated for a generic value of the U(1)$_R$ holonomy $\nu_R\in\mathbb{Z}/2$ \cite{Closset:2018ghr,Closset:2019hyt}, thereby extending the previous work restricted to the $A$-twist background with $\nu_R=0$ \cite{Closset:2017zgf}. For a non vanishing $\nu_R$, a vector multiplet contributes to the effective twisted superpoential in a non-trivial manner \cite{Closset:2018ghr,Closset:2019hyt}, but this contribution disappears in the parity-preserving quantization as\footnote{Here we have used $[-\alpha(\bu)]=1-[\alpha(\bu)]$, which is valid only if $\Re[\alpha(\bu)]\notin\mathbb{Z}$ that we assume throughout the paper. For ABJM theory of our interest, in particular, we confirm that the exact numerical ABJM Bethe vacua presented in \cite{Benini:2015eyy,Liu:2017vll,Bobev:2022eus} satisfy this assumption.}
	\begin{align}
		\fft{1}{(2\pi\ri)^2}\sum_{\alpha\in\mR[G]}\Li_2(\bx^{\alpha}e^{2\pi\ri(2\nu_R)})+\fft14\sum_{\alpha\in\mR[G]}B_2([\alpha(\bu)+2\nu_R])=0\,.
	\end{align}
	Here we have used the polylogarithm inversion formula 
	\begin{equation}
		\text{Li}_n(e^{2\pi\ri z})+(-1)^n\text{Li}_n(e^{-2\pi\ri z})=-\fft{(2\pi\ri)^n}{n!}\text{B}_n(z)~~~\begin{cases}
			0\leq\Re[z]<1~\&~\Im[z]\geq0 \\
			0<\Re[z]\leq1~\&~\Im[z]<0
		\end{cases}\label{polylog:inversion}
	\end{equation}
	given in terms of the Bernoulli polynomials, $2\nu_R\in\mathbb{Z}$, and that the contribution from a vector multiplet is equivalent to that from a chiral multiplet with $R$-charge $2$ \cite{Closset:2015rna,Closset:2018ghr}. This explains why (\ref{W}) does not include any non-trivial contributions from a vector multiplet. 
	
	\item Gauge and flavor flux operators are defined in terms of the effective twisted superpotential as
	\begin{equation}
		\Pi_i(\bu,\bnu)=\exp[2\pi\ri\fft{\partial\mW(\bu,\bnu)}{\partial u_i}]\quad\&\quad \Pi_I(\bu,\bnu)=\exp[2\pi\ri\fft{\partial\mW(\bu,\bnu)}{\partial\nu_I}]\label{Pi}
	\end{equation}
	respectively.
	
	\item Handle-gluing operators and the effective dilaton read
	\begin{align}
		\mH(\bu,\bnu)&=\exp[2\pi\ri\Omega(\bu,\bnu)]\det\mathbb{B}(\bu,\bnu)\,,\label{H}\\
		\Omega(\bu,\bnu)&=\sum_{i=1}^{|G|}k^{iR}u_i+\sum_{I=1}^{|H|}k^{IR}\nu_I+\fft12k^{RR}\nn\\
		&\quad-\fft{1}{2\pi\ri}\sum_{\alpha\in\mR[G]}\log(1-\bx^\alpha)+\fft{1}{2\pi\ri}\sum_\Psi(r_\Psi-1)\sum_{\rho_\Psi}\Li_1(\bx^{\rho_\Psi}\by^{\gamma_\Psi}e^{2\pi\ri r_\Psi\nu_R})\nn\\
		&\quad+\fft12\sum_\Psi(r_\Psi-1)\sum_{\rho_\Psi}B_1([\rho_\Psi(\bu)+\gamma_\Psi(\bnu)+r_\Psi\nu_R])\,,\label{Omega}\\
		\mathbb{B}(\bu,\bnu)&=\fft{\partial^2\mW(\bu,\bnu)}{\partial u_i\partial u_j}\,,\label{B}
	\end{align}
	where the last term in the effective dilaton (\ref{Omega}) is necessary for the parity-preserving quantization. Note that any shift in the effective dilaton by a real number --- used to determine the last term unambiguously in terms of the Bernoulli polynomial --- only affects the phase of the contribution to the $\mM_{\mg,p}$ partition function through the Bethe formula (\ref{Z}). Therefore, it does not influence $\Re\log Z_{\mM_{\mg,p}}$ of our interest, as long as we focus on the dominant contribution from a specific Bethe vacuum.

	\item Fibering operators are given by
	\begin{equation}
		\mF(\bu,\bnu)=\exp[2\pi\ri\bigg(\mW(\bu,\bnu)-\sum_{i=1}^{|G|}u_i\fft{\partial\mW(\bu,\bnu)}{\partial u_i}-\sum_{I=1}^{|H|}\nu_I\fft{\partial\mW(\bu,\bnu)}{\partial \nu_I}\bigg)]\,.\label{F}
	\end{equation}

	\item Bethe vacua are given by
	\begin{equation}
		\mS_{BE}=\Big\{\bu\,\bigg|\,\Pi_i(\bu,\bnu)=1,~w\cdot \bu\neq\bu~\text{for}~\forall w\in W\Big\}/W\,.\label{SBE}
	\end{equation}
\end{itemize}
%

\section{Supersymmetric partition functions of ABJM theory}\label{sec:ABJM}
In this section, we examine the $\mM_{\mg,p}$ partition function of the $\text{U}(N)_k\times\text{U}(N)_{-k}$ ABJM theory \cite{Aharony:2008ug}, utilizing the Bethe formulation introduced in the previous section \ref{sec:Mgp}. Subsection \ref{sec:ABJM:review} provides a brief overview of ABJM theory, with a focus on its global symmetries. In subsection \ref{sec:ABJM:Mgp-Bethe}, we present the $\mM_{\mg,p}$ partition function of ABJM theory in terms of the Bethe formulation. Finally, in subsection \ref{sec:ABJM:Mgp-eval}, we evaluate this partition function to all orders in the $1/N$-perturbative expansion by applying recent numerical analysis of the TTI \cite{Bobev:2022jte,Bobev:2022wem} to the Bethe formulation for the $\mM_{\mg,p}$ partition function.

\subsection{Lightning review of ABJM theory}\label{sec:ABJM:review}
The ABJM theory features the U$(N)_k\times$U$(N)_{-k}$ gauge group with opposite CS levels $\pm k$. Each unitary gauge group is associated with an $\mathcal N=2$ vector multiplet. The matter content consists of two pairs of bi-fundamental \& anti bi-fundamental $\mathcal N=2$ chiral multiplets, denoted as $A_{1,2}$ \& $B_{1,2}$ respectively, each with superconformal $R$-charge $\fft12$. The superpotential is given in terms of these chiral multiplets as
\begin{equation}
	W=\fft{4\pi}{k}\Tr[A_1B_1A_2B_2-A_1B_2A_2B_1]\,.
\end{equation}

Regarding the global symmetries, the ABJM theory exhibits the U$(1)_R$ superconformal $R$ symmetry and SU(2)$_A\times$SU(2)$_B\times$U(1)$_T$ flavor symmetries, as manifest in its $\mN=2$ formulation. The SU(2)$_A$ and SU(2)$_B$ flavor symmetries act on the multiplets $A_{1,2}$ and $B_{1,2}$ respectively, while U(1)$_T$ stands for the topological symmetry. Although there are two apparent U(1) topological symmetries, only the diagonal combination is non-trivial \cite{Aharony:2008ug,Freedman:2013oja,Bobev:2022wem,Bobev:2024mqw}. The Cartan generators of these global symmetries are denoted as $(R,A,B,T)$ respectively. The charges of the $\mathcal N=2$ chiral multiplets under these symmetries are summarized in Table~\ref{table:ABJM:global}, adapted from \cite{Aharony:2008ug,Benini:2015eyy}. 
\begin{table}
	\centering
	\begin{tabular}{|c||c|c|c|c|c|}
		\hline
		$\Phi$ & $R$ & $A$ & $B$ & $T$ & $b$ \\\hhline{|=||=|=|=|=|=|}
		$A_1$ & $\fft12$ & $1$ & $0$ & $0$ & $1$\\\hline
		$A_2$ & $\fft12$ & $-1$ & $0$ & $0$ & $1$\\\hline
		$B_1$ & $\fft12$ & $0$ & $1$ & $0$ & $-1$\\\hline
		$B_2$ & $\fft12$ & $0$ & $-1$ & $0$ & $-1$\\\hline
	\end{tabular}
	\caption{Charge assignments of $\mathcal N=2$ chiral multiplets in ABJM theory. The U$(1)_T$ symmetry acts non-trivially on monopole operators, which are omitted in the table.}
	\label{table:ABJM:global}
\end{table}

For simplicity, in the following calculations, we will treat the $U(1)_T$ topological symmetry as if it were replaced by the baryonic $U(1)_b$ symmetry with charges $b$ as shown in Table~\ref{table:ABJM:global}. This is in fact the case for the $\text{SU}(N)_k\times\text{SU}(N)_{-k}$ ABJM theory \cite{Aharony:2008ug}. The proper analysis in terms of the $U(1)_T$ topological symmetry for the $\text{U}(N)_k\times\text{U}(N)_{-k}$ ABJM theory can be reinstated by using an appropriate redefinition of gauge/flavor holonomies as well as magnetic fluxes. Such a redefinition has been explicitly applied to the matrix model associated with the SCI of ABJM theory \cite{Bobev:2022wem}. The procedure is quite parallel here, so we omit the details related to the necessary parameter redefinition for transitioning between the baryonic and topological symmetries.

\subsection{$\mM_{\mg,p}$ partition function of ABJM theory: Bethe formulation}\label{sec:ABJM:Mgp-Bethe}
In this subsection we present the Bethe formulation reviewed in subsection \ref{sec:Mgp:ptnfct} for ABJM theory. For compact expressions, it is useful to define the following parameters: 
\begin{subequations}
	\begin{align}
		\bDelta&=(\Delta_1,\Delta_2,\Delta_3,\Delta_4)&&=\fft{\nu_R}{2}I_4+(\nu_A+\nu_b,-\nu_A+\nu_b,\nu_B-\nu_b,-\nu_B-\nu_b)\,,\label{v:Delta}\\
		\bmn&=(\mn_1,\mn_2,\mn_3,\mn_4)&&=\fft{1-\mg}{2}I_4-(\mn_A+\mn_b,-\mn_A+\mn_b,\mn_B-\mn_b,-\mn_B-\mn_b)\,,\label{n:mn}
	\end{align}\label{ABJM:repara}%
\end{subequations}
where $I_4$ denotes the vector $I_4=(1,1,1,1)$. Based on the reparametrization (\ref{ABJM:repara}) and the charge assignments in Table~\ref{table:ABJM:global}, the Bethe formula for the $\mM_{\mg,p}$ partition function of ABJM theory, along with the corresponding building blocks, is expressed as follows. Note that $\bnu$ in the argument of the various building blocks is replaced with $\bDelta$ to account for the reparametrization given in (\ref{ABJM:repara}).
\begin{itemize}
	\item The effective twisted superpotential for ABJM theory from (\ref{W}): 
	\begin{align}
		\mW^\text{ABJM}(\bu,\bDelta)&=\fft{k}{2}\sum_{i=1}^N(u_{1,i}(u_{1,i}+1+2\nu_R)-u_{2,i}(u_{2,i}+1+2\nu_R))\label{ABJM:W}\\
		&\quad+\sum_{i,j=1}^N\sum_{a=1}^4\Bigg[\fft{1}{(2\pi\ri)^2}\Li_2(e^{2\pi\ri(\sigma_a(u_{1,i}-u_{2,j})+\Delta_a)})+\fft14B_2([\sigma_a(u_{1,i}-u_{2,j})+\Delta_a])\Bigg]\,,\nn
	\end{align}
	where we have introduced $\sigma_a=(1,1,-1,-1)$.

	\item ABJM gauge and flavor flux operators from (\ref{Pi}):\footnote{We have implicitly assumed that $\Re[\sigma_a(u_{1,i}-u_{2,j})+\Delta_a]$ do not take on integer values in order to ensure the well-defined first derivatives of the effective twisted superpotential.}
	\begin{align}
		&\Pi^\text{ABJM}_{r,i}(\bu,\bDelta)=\exp[2\pi\ri\fft{\partial\mW^\text{ABJM}(\bu,\bDelta)}{\partial u_{r,i}}]\\
		&=\begin{cases}
			(-1)^{k(1+2\nu_R)}x_{1,i}^k\prod_{a=1}^4\prod_{j=1}^N\Big(\fft{e^{\pi\ri(\sigma_a(u_{1,i}-u_{2,j})+\Delta_a)}e^{\pi\ri\varphi_a}}{1-e^{2\pi\ri(\sigma_a(u_{1,i}-u_{2,j})+\Delta_a)}}\Big)^{\sigma_a} & (r=1) \\[0.5em]
			(-1)^{-k(1+2\nu_R)}x_{2,j}^{-k}\prod_{a=1}^4\prod_{j=1}^N\Big(\fft{e^{\pi\ri(\sigma_a(u_{1,i}-u_{2,j})+\Delta_a)}e^{\pi\ri\varphi_a}}{1-e^{2\pi\ri(\sigma_a(u_{1,i}-u_{2,j})+\Delta_a)}}\Big)^{-\sigma_a} & (r=2)
		\end{cases}\,,\nn
	\end{align}
	\begin{align}
		\Pi^\text{ABJM}_I(\bu,\bDelta)&=\exp[2\pi\ri\fft{\partial\mW^\text{ABJM}(\bu,\bDelta)}{\partial \nu_I}]\\
		&=\begin{cases}
			\prod_{i,j=1}^N\Big(\fft{e^{\pi\ri(u_{1,i}-u_{2,j}+\Delta_1)}e^{\pi\ri\varphi_1}}{1-e^{2\pi\ri(u_{1,i}-u_{2,j}+\Delta_1)}}\Big)\Big(\fft{e^{\pi\ri(u_{1,i}-u_{2,j}+\Delta_2)}e^{\pi\ri\varphi_2}}{1-e^{2\pi\ri(u_{1,i}-u_{2,j}+\Delta_2)}}\Big)^{-1} & (I=A) \\[0.5em]
			\prod_{i,j=1}^N\Big(\fft{e^{\pi\ri(-u_{1,i}+u_{2,j}+\Delta_3)}e^{\pi\ri\varphi_3}}{1-e^{2\pi\ri(-u_{1,i}+u_{2,j}+\Delta_3)}}\Big)\Big(\fft{e^{\pi\ri(-u_{1,i}+u_{2,j}+\Delta_4)}e^{\pi\ri\varphi_4}}{1-e^{2\pi\ri(-u_{1,i}+u_{2,j}+\Delta_4)}}\Big)^{-1} & (I=B) \\[0.5em]
			\prod_{a=1}^4\prod_{i,j=1}^N\Big(\fft{e^{\pi\ri(\sigma_a(u_{1,i}-u_{2,j})+\Delta_a)}e^{\pi\ri\varphi_a}}{1-e^{2\pi\ri(\sigma_a(u_{1,i}-u_{2,j})+\Delta_a)}}\Big)^{\sigma_a} & (I=b) \\
		\end{cases}\,.\nn
	\end{align}
	Considering the reparametrization (\ref{ABJM:repara}), it is useful to introduce a new flavor flux operator with respect to $\Delta_a$ as
	\begin{align}
		\Pi^\text{ABJM}_a(\bu,\bDelta)=\exp[2\pi\ri\fft{\partial\mW^\text{ABJM}(\bu,\bDelta)}{\partial\Delta_a}]=\prod_{i,j=1}^N\fft{e^{\pi\ri(\sigma_a(u_{1,i}-u_{2,j})+\Delta_a)}e^{\pi\ri\varphi_a}}{1-e^{2\pi\ri(\sigma_a(u_{1,i}-u_{2,j})+\Delta_a)}}\,.
	\end{align}
	It is then straightforward to derive
	\begin{align}
		\prod_{I\in\{A,B,b\}}\Pi^\text{ABJM}_I(\bu,\bDelta)^{\mn_I}=\prod_{a=1}^4\Pi_a^\text{ABJM}(\bu,\bDelta)^{\fft{1-\mg}{2}-\mn_a}\,.\label{I:to:a}
	\end{align}
	Note that we have introduced constant phases $e^{\pi\ri\varphi_a}~(\varphi_a\in\mathbb{R})$ in the above flux operators that arise from $e^{\pi\ri B_1([\sigma_a(u_{1,i}-u_{2,j})+\Delta_a])}=e^{\pi\ri(\sigma_a(u_{1,i}-u_{2,j})+\Delta_a)}e^{\pi\ri\varphi_a}$, which are multiples of $\ri$ by construction. These phases in the flavor flux operators will not affect $\Re\log Z^\text{ABJM}_{\mM_{\mg,p}}$ evaluated through the Bethe formula (\ref{ABJM:Z}) at the end of the day. The same phases in the gauge flux operator, on the other hand, must be carefully tracked to accurately determine the Bethe vacua, which will be addressed in the next subsection \ref{sec:ABJM:Mgp-eval}. 
	
	\item ABJM handle-gluing operators from (\ref{H}): 
	\begin{align}
		\mH^\text{ABJM}(\bu,\bDelta)&=\exp[2\pi\ri\Omega^\text{ABJM}(\bu,\bDelta)]\det\mathbb{B}^\text{ABJM}(\bu,\bDelta)\,,\\
		\Omega^\text{ABJM}(\bu,\bDelta)&=-\fft{1}{2\pi\ri}\sum_{i\neq j}^N\bigg[\log(1-\fft{x_{1,i}}{x_{1,j}})+\log(1-\fft{x_{2,i}}{x_{2,j}})\bigg]\nn\\
		&\quad+\fft{1}{2\pi\ri}\sum_{i,j=1}^N(\fft12-1)\sum_{a=1}^4\Li_1(e^{2\pi\ri(\sigma_a(u_{1,i}-u_{2,j})+\Delta_a)})\nn\\
		&\quad+\fft12\sum_{i,j=1}^N(\fft12-1)\sum_{a=1}^4B_1([\sigma_a(u_{1,i}-u_{2,j})+\Delta_a])\,,\label{ABJM:Omega}\\
		\mathbb{B}^\text{ABJM}(\bu,\bDelta)&=\fft{\partial^2\mW^\text{ABJM}(\bu,\bDelta)}{\partial u_{r,i}\partial u_{s,j}}\qquad(r,s\in\{1,2\})\,.
	\end{align}

	\item ABJM fibering operators from (\ref{F}): 
	\begin{align}
		&\mF^\text{ABJM}(\bu,\bDelta)=\exp\Bigg[2\pi\ri\bigg(1-\sum_{r=1}^2\sum_{i=1}^{N}u_{r,i}\fft{\partial}{\partial u_{r,i}}-\sum_{I\in\{A,B,b\}}\nu_I\fft{\partial}{\partial \nu_I}\bigg)\mW^\text{ABJM}(\bu,\bDelta)\Bigg]\nn\\
		&=\exp\Bigg[2\pi\ri\bigg(1-\sum_{r=1}^2\sum_{i=1}^{N}u_{r,i}\fft{\partial}{\partial u_{r,i}}-\sum_{a=1}^4\Big(\Delta_a-\fft{\sum_{b=1}^4\Delta_b}{4}\Big)\fft{\partial}{\partial\Delta_a}\bigg)\mW^\text{ABJM}(\bu,\bDelta)\Bigg]\,.\label{ABJM:F}
	\end{align}

	\item ABJM Bethe vacua from (\ref{SBE}):
	\begin{equation}
		\begin{split}
			\mS_{BE}^\text{ABJM}&=\bigg\{\bu\,\bigg|\,\Pi^\text{ABJM}_{r,i}(\bu,\bDelta)=1,~w\cdot \bu\neq\bu~\text{for}~\forall w\in W\bigg\}/W\,.
		\end{split}\label{ABJM:SBE}
	\end{equation}
\end{itemize}
In terms of the building blocks outlined above, the Bethe formula for the $\mM_{\mg,p}$ partition function of ABJM theory reads
\begin{align}
	Z^\text{ABJM}_{\mM_{\mg,p}}(\bDelta,\bmn)=\sum_{\bu\in\mS^\text{ABJM}_{BE}}\mF^\text{ABJM}(\bu,\bDelta)^p\mH^\text{ABJM}(\bu,\bDelta)^{\mg-1}\prod_{a=1}^4\Pi_a^\text{ABJM}(\bu,\bDelta)^{\fft{1-\mg}{2}-\mn_a}\,.\label{ABJM:Z}
\end{align}
Focusing on the contribution from a particular Bethe vacuum $\{\bu_\star=\bu_\star(\bDelta)\}$, the Bethe formula (\ref{ABJM:Z}) can be recast into
\begin{align}
	\log Z^\text{ABJM}_{\mM_{\mg,p}}(\bDelta,\bmn)&=p\log\mF^\text{ABJM}(\bu_\star,\bDelta)+\log Z^\text{ABJM}_{\mM_{\mg,0}}(\bDelta,\bmn)\,,\label{ABJM:logZ}\\
	\log Z^\text{ABJM}_{\mM_{\mg,0}}(\bDelta,\bmn)&=(\mg-1)\log\det\mathbb{B}(\bu_\star,\bDelta)+(1-\mg)\sum_{i\neq j}^N\bigg[\log(1-\fft{x_{\star\,1,i}}{x_{\star\,1,j}})+\log(1-\fft{x_{\star\,2,i}}{x_{\star\,2,j}})\bigg]\nn\\
	&\quad+\sum_{a=1}^4(1-\mg-\mn_a)\sum_{i,j=1}^N\log\bigg(\fft{e^{\pi\ri(\sigma_a(u_{\star\,1,i}-u_{\star\,2,j})+\Delta_a)}e^{\ri\varphi_a}}{1-e^{2\pi\ri(\sigma_a(u_{\star\,1,i}-u_{\star\,2,j})+\Delta_a)}}\bigg)\nn\\
	&\quad+(\text{contribution from other Bethe vacua})\,,\nn
\end{align}
where we have used more explicit expressions for the building blocks. Note that the Bethe vacuum $\bu_\star(\bDelta)$ depends on the given parameters $\bDelta$ but for notational convenience we often omit this argument. 

\medskip

In the next subsection \ref{sec:ABJM:Mgp-eval}, we will omit the contribution from other Bethe vacua. As we will explicitly verify in section \ref{sec:holo}, the contribution from the specific Bethe vacuum $\{\bu_\star\}$, which will be identified in the following subsection, precisely matches the dual quantity of the Euclidean AdS-Taub-Bolt geometry examined in section \ref{sec:sugra} through the AdS/CCFT correspondence, without incorporating contributions from other Bethe vacua. However, exploring the existence of additional Bethe vacua and their corresponding contributions to the partition function through the Bethe formula is in itself a crucial subject in this line of research for a comprehensive picture, which may also shed light on new dual supergravity backgrounds sharing the same asymptotic structure as the Euclidean AdS-Taub-Bolt geometry. We will comment on related aspects in section \ref{sec:discussion}.

\subsection{$\mM_{\mg,p}$ partition function of ABJM theory: evaluation}\label{sec:ABJM:Mgp-eval}
To evaluate the $\mM_{\mg,p}$ partition function of ABJM theory using the Bethe formula (\ref{ABJM:logZ}), the first step is to identify the Bethe vacuum $\{\bu_\star\}$. Rather than solving the Bethe equations (\ref{ABJM:SBE}) from scratch, we can employ the established solutions to the Bethe-Ansatz-Equations (BAE) for the $S^1\times\Sigma_{\mg}$ TTI of ABJM theory. These BAE solutions were originally derived analytically in the large $N$ limit \cite{Benini:2015eyy} and later refined for finite $N$ through numerical analysis \cite{Liu:2017vll,Bobev:2022jte,Bobev:2022eus}. To demonstrate the equivalence between the BAE solutions and the Bethe vacua explicitly, we first verify that the BAE for the TTI of ABJM theory presented in \cite{Bobev:2022eus,Bobev:2022wem} correspond to the following equations\footnote{For a detailed comparison, map the parameters as
	\begin{align}
		u_i^\text{\cite{Bobev:2022eus,Bobev:2022wem}}=2\pi u_{1,i}\,,\qquad \tu_i^\text{\cite{Bobev:2022eus,Bobev:2022wem}}=2\pi u_{2,i}\,,\qquad \Delta_a^\text{\cite{Bobev:2022eus,Bobev:2022wem}}=2[\Delta_a]\,.\label{convention:matching}
	\end{align}
}
\begin{subequations}
	\begin{align}
		\fft{\partial\mW^\text{ABJM}(\bu,\bDelta)}{\partial u_{1,i}}&=\fft{k(1+2\nu_R)+(N-2i+2)}{2}-\fft{1-(-1)^N}{4}\,,\\
		\fft{\partial\mW^\text{ABJM}(\bu,\bDelta)}{\partial u_{2,i}}&=-\fft{k(1+2\nu_R)-(N-2i)}{2}+\fft{1-(-1)^N}{4}\,,
	\end{align}\label{ABJM:SBE:explicit}%
\end{subequations}
where we have focused on the case with\footnote{Under the assumption (\ref{Stokes}), there are three possible values for the sum of $[\Delta_a]$, namely $\sum_{a=1}^4[\Delta_a]\in\{1,2,3\}$. We will first consider the case where $\sum_{a=1}^4[\Delta_a]=1$ and subsequently address the case where $\sum_{a=1}^4[\Delta_a]=3$ using the symmetry (\ref{ABJM:Z:symmetry}) below. There is no known Bethe vacuum for the case where $\sum_{a=1}^4[\Delta_a]=2$, thus we will not discuss it further in this manuscript, see \cite{Toldo:2017qsh} for related comments.}
\begin{equation}
	\sum_{a=1}^4\Delta_a=2\nu_R\in\mathbb{Z}\qquad\to\qquad \sum_{a=1}^4[\Delta_a]=1\,,\label{sum:Delta}
\end{equation}
and further assumed
\begin{equation}
	[\sigma_a(u_i-\tu_j)+\Delta_a]=\sigma_a(u_i-\tu_j)+[\Delta_a]\,,
\end{equation}
for the application of the polylogarithm inversion formula (\ref{polylog:inversion}). Importantly, we note that the right-hand sides of (\ref{ABJM:SBE:explicit}) yield integers for half-integer values of $\nu_R$ and integer values of $(N,k)$.\footnote{In the parity-preserving quantization, the CS levels need to be appropriately shifted in terms of the weights of the chiral multiplets within a given theory to ensure that they take on integer values \cite{Aharony:1997bx,Bobev:2024mqw}. In the case of ABJM theory under consideration, this shift itself results in an integer value, allowing us to impose the integer condition on the CS level $k$.} Although $\nu_R$ can be either an integer or a half-integer as discussed below (\ref{W}), the dual supergravity analysis in the subsequent sections \ref{sec:sugra} and \ref{sec:holo} will constrain $\nu_R$ to $\nu_R\in\{\pm\fft12\}$, which is crucial for ensuring that the right-hand sides of (\ref{ABJM:SBE:explicit}) are integer values. This concludes that the BAEs for the TTI, as represented in (\ref{ABJM:SBE:explicit}) under the current conventions, correspond precisely to the Bethe equations in (\ref{ABJM:SBE}) provided we focus on the $\mM_{\mg,p}$ partition function with the appropriate supergravity dual. Hence we can confidently use the exact BAE solution constructed in \cite{Bobev:2022eus} as the Bethe vacuum $\{\bu_\star\}$ that solves the Bethe equations.

\medskip

Thanks to the equivalence between the BAE solution in \cite{Bobev:2022eus} and the Bethe vacuum $\{\bu_\star\}$, the second term $\log Z^\text{ABJM}_{\mM_{\mg,0}}$ in the Bethe formula (\ref{ABJM:logZ}) aligns perfectly with the logarithm of the $S^1\times\Sigma_{\mg}$ TTI studied in \cite{Bobev:2022eus}. To be more precise, this equivalence holds for real $\bDelta$, allowing us to disregard negligible imaginary terms defined modulo $2\pi\ri\mathbb{Z}$, which arise from the inherent ambiguity in the choice of logarithm branches. As a result, assuming real $\bDelta$ henceforth, we obtain
\begin{align}
	\Re\log Z^\text{ABJM}_{\mM_{\mg,0}}&=-\fft{2\pi\sqrt{2k\prod_{a=1}^4[\Delta_a]}}{3}\sum_{a=1}^4\fft{\mn_a}{[\Delta_a]}\bigg(\hN_{k,\bDelta}^{3/2}-\fft{\mfc_a}{k}\hN_{k,\bDelta}^{1/2}\bigg)-\fft{1-\mg}{2}\log\hN_{k,\bDelta}\nn\\
	&\quad+\hf_0(k,\bDelta,\bmn)+\mO(e^{-\sqrt{N}})\label{ABJM:logZ:p=0}
\end{align}
from the closed-from expression for the $S^1\times\Sigma_{\mg}$ TTI deduced in \cite{Bobev:2022eus}. It is expressed in terms of the following parameters
\begin{subequations}
\begin{align}
	\hN_{k,\bDelta}&=N-\fft{k}{24}+\fft{1}{24k}\sum_{a=1}^4\fft{1}{[\Delta_a]}\,,\\
	\mfc_a&=\fft{\prod_{b=1\,(\neq a)}^4([\Delta_a]+[\Delta_b])}{8\prod_{b=1}^4[\Delta_b]}\sum_{b=1\,(\neq a)}^4[\Delta_b]\,.
\end{align}\label{hN:mfc}%
\end{subequations}
In (\ref{ABJM:logZ:p=0}), the term $\hf(k,\bDelta,\bmn)$ represents an $N$-independent constant whose numerical values are presented for various configurations of $(k,\bDelta,\bmn)$ in \cite{Bobev:2022eus}. The notation $\mO(e^{-\sqrt{N}})$ encapsulates exponentially suppressed non-perturbative corrections.

\medskip

The next step for the evaluation of the $\mM_{\mg,p}$ partition function of ABJM theory via the Bethe formula is to analyze the contribution from the fibering operator $p\log\mF^\text{ABJM}(\bu_\star,\bDelta)$, the first term in the Bethe formula (\ref{ABJM:logZ}). Since the fibering operator (\ref{ABJM:F}) is written in terms of the effective twisted superpotential (\ref{ABJM:W}), we begin by determining its on-shell value at the Bethe vacuum $\{\bu_\star\}$ as
\begin{align}
	\mW^\text{ABJM}(\bu_\star,\bDelta)&=\fft{1}{2\pi\ri}\Bigg[\fft{4\pi\sqrt{2k\prod_{a=1}^4[\Delta_a]}}{3}\hN_{k,\bDelta}^{3/2}+\hg_0(k,\bDelta)+\mO(e^{-\sqrt{N}})\Bigg]+\mR\label{ABJM:W:on-shell}\\
	&\quad+\sum_{i=1}^N\bigg(\fft{k(1+2\nu_R)+(N-2i+2)}{2}-\fft{1-(-1)^N}{4}\bigg)u_{\star\,1,i}\nn\\
	&\quad+\sum_{i=1}^N\bigg(-\fft{k(1+2\nu_R)-(N-2i)}{2}+\fft{1-(-1)^N}{4}\bigg)u_{\star\,2,i}\,.\nn
\end{align}
This expression is derived by employing the closed-form expression for the on-shell Bethe potential at the BAE solution corresponding to the Bethe vacuum $\{\bu_\star\}$ \cite{Bobev:2022eus}.\footnote{For a detailed comparison, we have employed
\begin{align}
	\mW^\text{ABJM}&=\fft{1}{(2\pi\ri)^2}\mV_\text{TTI}^\text{\cite{Bobev:2022wem}}+\sum_{i=1}^N\bigg(\fft{k(1+2\nu_R)+(N-2i+2)}{2}-\fft{1-(-1)^N}{4}\bigg)u_{1,i}\nn\\
	&\quad+\sum_{i=1}^N\bigg(-\fft{k(1+2\nu_R)-(N-2i)}{2}+\fft{1-(-1)^N}{4}\bigg)u_{2,i}+(\bu\text{-independent real terms})\nn
\end{align}
under the parameter matching (\ref{convention:matching}).} In (\ref{ABJM:W:on-shell}), $\mR$ encompasses the insignificant real parts that do not influence the value of $\Re\log Z^\text{ABJM}_{\mg,p}$ of interest, and $\hg(k,\bDelta)$ is an $N$-independent constant whose numerical values are presented for various $(k,\bDelta)$ configurations in \cite{Bobev:2022wem}. Substituting the on-shell effective twisted superpotential (\ref{ABJM:W:on-shell}) and the Bethe equations (\ref{ABJM:SBE:explicit}) into the ABJM fibering operator (\ref{ABJM:F}), we obtain its on-shell value as
\begin{align}
	\Re\log\mF^\text{ABJM}(\bu_\star,\bDelta)&=\bigg(1-\sum_{a=1}^4\Big(\Delta_a-\fft{\sum_{b=1}^4\Delta_b}{4}\Big)\fft{\partial}{\partial\Delta_a}\bigg)\Bigg[\fft{4\pi\sqrt{2k\prod_{a=1}^4[\Delta_a]}}{3}\hN_{k,\bDelta}^{3/2}+\hg_0(k,\bDelta)\Bigg]\nn\\
	&\quad+\mO(e^{-\sqrt{N}})\,.\label{ABJM:logF:1}
\end{align}
To circumvent subtleties related to Stokes' lines (as discussed in \cite{Benini:2018ywd} in a different context), we assume
\begin{equation}
	\Delta_a\notin\mathbb{Z}\quad\to\quad\fft{\partial}{\partial\Delta_a}=\fft{\partial}{\partial[\Delta_a]}\,.\label{Stokes}
\end{equation}
The on-shell ABJM fibering operator (\ref{ABJM:logF:1}) can then be expressed more explicitly as
\begin{align}
	\Re\log\mF^\text{ABJM}(\bu_\star,\bDelta)&=\fft{2\pi\sqrt{2k[\Delta_1][\Delta_2][\Delta_3][\Delta_4]}}{3}\hN_{k,\bDelta}^{3/2}\bigg(2-\sum_{a=1}^4\fft{\Delta_a-\fft{\sum_{b=1}^4\Delta_b}{4}}{[\Delta_a]}\bigg)\nn\\
	&\quad+\fft{\pi\sqrt{2k[\Delta_1][\Delta_2][\Delta_3][\Delta_4]}}{12k}\hN_{k,\bDelta}^{1/2}\sum_{a=1}^4\fft{\Delta_a-\fft{\sum_{b=1}^4\Delta_b}{4}}{[\Delta_a]^2}\nn\\
	&\quad+\bigg(1-\sum_{a=1}^4\Big(\Delta_a-\fft{\sum_{b=1}^4\Delta_b}{4}\Big)\fft{\partial}{\partial\Delta_a}\bigg)\hat{g}_0(k,\bDelta)+\mO(e^{-\sqrt{N}})\,. \label{ABJM:logF:2}
\end{align}

\medskip

Finally, substituting the $\mM_{\mg,0}$ partition function (\ref{ABJM:logZ:p=0}) and the fibering operator (\ref{ABJM:logF:2}) into the Bethe formula (\ref{ABJM:logZ}), we arrive at the closed-form expression for the $\mM_{\mg,p}$ partition function valid to all orders in the $1/N$-perturbative expansion:
\begin{align}
	&\Re\log Z^\text{ABJM}_{\mM_{\mg,p}}(\bDelta,\bmn)\label{ABJM:logZ:result}\\
	&=p\fft{2\pi\sqrt{2k\prod_{a=1}^4[\Delta_a]}}{3}\Bigg[\hat N_{k,\bDelta}^{3/2}\bigg(2-\sum_{a=1}^4\fft{\Delta_a-\fft{\sum_{b=1}^4\Delta_b}{4}}{[\Delta_a]}\bigg)+\fft{1}{8k}\hat N_{k,\bDelta}^{1/2}\sum_{a=1}^4\fft{\Delta_a-\fft{\sum_{b=1}^4\Delta_b}{4}}{[\Delta_a]^2}\Bigg]\nn\\
	&\quad-\fft{2\pi\sqrt{2k\prod_{a=1}^4[\Delta_a]}}{3}\sum_{a=1}^4\fft{\mn_a}{[\Delta_a]}\bigg(\hN_{k,\bDelta}^{3/2}-\fft{\mfc_a}{k}\hN_{k,\bDelta}^{1/2}\bigg)-\fft{1-\mg}{2}\log\hN_{k,\bDelta}\nn\\
	&\quad+p\bigg(1-\sum_{a=1}^4\Big(\Delta_a-\fft{\sum_{b=1}^4\Delta_b}{4}\Big)\fft{\partial}{\partial\Delta_a}\bigg)\hat{g}_0(k,\bDelta)+\hf_0(k,\bDelta,\bmn)+\mO(e^{-\sqrt{N}})\,.\nn
\end{align}
Recall that we have utilized expressions introduced in (\ref{ABJM:repara}) and (\ref{hN:mfc}) satisfying the constraints
\begin{equation}
	\sum_{a=1}^4[\Delta_a]=1\,,\qquad\sum_{a=1}^4\mn_a=2(1-\mg)\,.\label{constraint}
\end{equation}
Below we present several comments on the closed-form expression (\ref{ABJM:logZ:result}).
\begin{itemize}
	\item The fist two terms in the last line of (\ref{ABJM:logZ:result}) are independent of the expansion parameter $N$. While closed-form expressions for these terms are not yet available, their numerical values can be found in \cite{Bobev:2022eus,Bobev:2022wem} for various configurations $(k,\bDelta,\bmn)$. The final term, of order $\mO(e^{-\sqrt{N}})$, captures the non-perturbative corrections.
	
	\item In the strict large $N$ limit, the all order $\mM_{\mg,p}$ partition function (\ref{ABJM:logZ:result}) reduces to 
	\begin{align}
		&\Re\log Z^\text{ABJM}_{\mM_{\mg,p}}(\bDelta,\bmn)\label{ABJM:logZ:largeN}\\
		&=-\fft{2\pi\sqrt{2k\prod_{a=1}^4[\Delta_a]}}{3}\bigg[2p+\sum_{a=1}^4\fft{p(D_a-\fft{\sum_{b=1}^4\Delta_b}{4})+\mn_a}{[\Delta_a]}\bigg]N^{3/2}+\mO(N^{1/2})\,,\nn
	\end{align}
	where we define an integer $D_a$ such that $\Delta_a=[\Delta_a]+D_a$. This is consistent with the previous large $N$ analysis presented in \cite{Toldo:2017qsh} under the convention matching
	\begin{equation}
		(m_i,n_i,-s_i-(\mg-1)r_i)^\text{\cite{Toldo:2017qsh}}\quad\leftrightarrow\quad(\Delta_a,D_a,\mn_a)\nn
	\end{equation}
	and the appropriate choice of holonomy for the $R$ symmetry background gauge field, which constrains the sum of the $\bDelta$ parameters by $\sum_{b=1}^4\Delta_b=2\nu_R$. Our result (\ref{ABJM:logZ:result}) significantly enhances the previous work, extending it from the $N^{3/2}$ leading order to all orders in the $1/N$-perturbative expansion up to exponentially suppressed non-perturbative corrections.
	
	\item The large gauge transformation along the Seifert fiber requires any physical observable remains invariant under the transformation \cite{Closset:2017zgf}
	\begin{equation}
		(\Delta_a,\mn_a)\quad\to\quad(\Delta_a+1,\mn_a-p)\,,\label{sym}
	\end{equation}
	where the opposite signs for the shifts stem from the reparametrization (\ref{ABJM:repara}). The $\mM_{\mg,p}$ partition function (\ref{ABJM:logZ:result}) is not invariant under the large gauge transformation (\ref{sym}), however, since it violates the second constraint in (\ref{constraint}) necessary for the validity of expression (\ref{ABJM:logZ:result}). Instead, we find that the $\mM_{\mg,p}$ partition function (\ref{ABJM:logZ:result}) is invariant under
	\begin{equation}
		(\Delta_a,\mn_a,\Delta_b,\mn_b)\to(\Delta_a+1,\mn_a-p,\Delta_b-1,\mn_b+p)\label{sym:2}
	\end{equation}
	for any distinct $a,b\in\{1,2,3,4\}$, which preserves both constraints in (\ref{constraint}). Closed-form expressions for the $N$-independent terms have not yet been established, leaving their invariance under the large gauge transformation to be explored in future research. At least in the large $k$ limit, however, the $k^2$ leading orders of $\hg_0(k,\bDelta)$ and $\hf_0(k,\bDelta,\bmn)$ have recently been determined as \cite{Bobev:2022eus,Hong:2024}\footnote{The coefficients $\hf_{0,2,a}(\bDelta)$ with $a\in\{2,3,4\}$ can be derived by employing the symmetries \cite{Bobev:2022eus}
	\begin{equation}
		\hf_{0,2,2}(\bDelta)=\hf_{0,2,1}(\bDelta)|_{[\Delta_1]\leftrightarrow[\Delta_2]} \, , \quad
		\hf_{0,2,3}(\bDelta)=\hf_{0,2,1}(\bDelta)|_{([\Delta_1],[\Delta_2])\leftrightarrow([\Delta_3],[\Delta_4])}=
		\hf_{0,2,4}(\bDelta)|_{[\Delta_3]\leftrightarrow[\Delta_4]}\,.
	\end{equation}
	}
	\begin{subequations}
	\begin{align}
		\hg_0(k,\bDelta)&=\bigg[1-\sum_{a=1}^4[\Delta_a]^2-2\bigg(\fft{1}{[\Delta_{13}][\Delta_{24}]}+\fft{1}{[\Delta_{14}][\Delta_{23}]}\bigg)\sum_{a=1}^4\fft{\prod_{b=1}^4[\Delta_b]}{[\Delta_a]}\bigg]\fft{\zeta(3)}{8\pi^2}k^2\nn\\
		&\quad+o(k^2)\,,\\[0.5em]
		\hf_0(k,\bDelta,\bmn)&=-\bigg[\sum_{a=1}^4\hf_{0,2,a}(\bDelta)\mn_a\bigg]\fft{\zeta(3)}{4\pi^2}k^2+o(k^2)\,,\\
		\hf_{0,2,1}(\bDelta)&=[\Delta_1]+\fft{[\Delta_1][\Delta_3]}{[\Delta_{14}]}+\fft{[\Delta_1][\Delta_4]}{[\Delta_{13}]}+\fft{[\Delta_1][\Delta_4][\Delta_{23}]}{[\Delta_{14}]^2}+\fft{[\Delta_1][\Delta_3][\Delta_{24}]}{[\Delta_{13}]^2}\nn\\
		&\quad-\fft{[\Delta_3][\Delta_4]}{[\Delta_{13}][\Delta_{14}]}-\fft{[\Delta_2]^2([\Delta_1]-[\Delta_2])}{[\Delta_{23}][\Delta_{24}]}+\fft{[\Delta_2][\Delta_3][\Delta_{14}]}{[\Delta_{13}][\Delta_{23}]}+\fft{[\Delta_2][\Delta_4][\Delta_{13}]}{[\Delta_{14}][\Delta_{24}]}\,.\nn
	\end{align}\label{hg:hf}%
	\end{subequations}
	Using the expressions (\ref{hg:hf}), we confirm that the $N$ independent terms in (\ref{ABJM:logZ:result}) are indeed invariant under the large gauge transformation (\ref{sym:2}) at the $k^2$ leading order. Here we have used the shorthand notation $[\Delta_{ab}]=[\Delta_a]+[\Delta_b]$. This serves as a highly non-trivial consistency check for the expressions (\ref{hg:hf}) determined via numerical analysis in \cite{Bobev:2022eus,Hong:2024}.
	
	\item The $\mM_{\mg,p}$ partition function for the case where $\sum_{a=1}^4[\Delta_a]=3$ can be easily derived by using the symmetry
	\begin{equation}
		Z^\text{ABJM}_{\mM_{\mg,p}}(\bDelta,\bmn)=Z^\text{ABJM}_{\mM_{\mg,-p}}(-\bDelta,\bmn)\,,\label{ABJM:Z:symmetry}
	\end{equation}
	generalizing a similar observation made in \cite{Benini:2015eyy} for non-zero $p$. To be more specific, leveraging the symmetry (\ref{ABJM:Z:symmetry}) and the relation
	\begin{equation}
		\sum_{a=1}^4[\Delta_a]=3\quad\xrightarrow{[\Delta_a]=1-[-\Delta_a]}\quad\sum_{a=1}^4[-\Delta_a]=1\,,
	\end{equation}
	we can obtain the $\mM_{\mg,p}$ partition function for $\sum_{a=1}^4[\Delta_a]=3$ by simply replacing $(\Delta_a,[\Delta_a],p)$ with $(-\Delta_a,1-[\Delta_a],-p)$ in the expression (\ref{ABJM:logZ:result}), where we have implicitly assumed (\ref{Stokes}) for the identity $[\Delta_a]=1-[-\Delta_a]$. 
	
	To prove the symmetry (\ref{ABJM:Z:symmetry}), we first derived the relation
	\begin{align}
		\mW^\text{ABJM}(\bu,-\bDelta)&=-\mW^\text{ABJM}(\bu,\bDelta)\Big|_{u_{1,i}\leftrightarrow u_{2,i}}
	\end{align}
	for the effective twisted superpotential (\ref{ABJM:W}). Consequently, the Bethe vacuum (\ref{ABJM:SBE}) transforms under the sign flip of the $\bDelta$ parameters as 
	\begin{equation}
	\begin{split}
		\Pi^\text{ABJM}_{1,i}(\bu,-\bDelta)&=\Pi^\text{ABJM}_{2,i}(\bu,\bDelta)^{-1}\Big|_{u_{1,i}\leftrightarrow u_{2,i}}\\
		\Pi^\text{ABJM}_{2,i}(\bu,-\bDelta)&=\Pi^\text{ABJM}_{1,i}(\bu,\bDelta)^{-1}\Big|_{u_{1,i}\leftrightarrow u_{2,i}}
	\end{split}\quad\to\quad\bu_\star(-\bDelta)=\bu_\star(\bDelta)\Big|_{u_{1,i}\leftrightarrow u_{2,i}}\,,
	\end{equation}
	where we have reinstated the argument of the Bethe vacuum to clarify its dependence on the $\bDelta$ parameters. The on-shell values of the building blocks for the Bethe formula presented in subsection \ref{sec:ABJM:Mgp-Bethe} are then given for the sign flipped $\bDelta$ parameters as
	\begin{subequations}
	\begin{align}
		\Pi^\text{ABJM}_a(\bu_\star(-\bDelta),-\bDelta)&=\Pi^\text{ABJM}_a(\bu_\star(\bDelta),\bDelta)\,,\\
		\det\mathbb{B}^\text{ABJM}(\bu_\star(-\bDelta),-\bDelta)&=\det\mathbb{B}^\text{ABJM}(\bu_\star(\bDelta),\bDelta)\,,\\
		\Omega^\text{ABJM}(\bu_\star(-\bDelta),-\bDelta)&=\Omega^\text{ABJM}(\bu_\star(\bDelta),\bDelta)\,,\\
		\mF^\text{ABJM}(\bu_\star(-\bDelta),-\bDelta)&=\mF^\text{ABJM}(\bu_\star(\bDelta),\bDelta)^{-1}\,.
	\end{align}
	\end{subequations}
	Substituting these expressions back into the Bethe formula (\ref{ABJM:Z}) then demonstrates the symmetry (\ref{ABJM:Z:symmetry}).
\end{itemize}
%

\section{Supergravity dual}\label{sec:sugra}
In this section, we review the supergravity background holographically dual to the $\mM_{\mg,p}$ partition function of ABJM theory discussed in section \ref{sec:ABJM}, focusing on the universal twist wherein the background gauge field is turned on only for the superconformal $R$ symmetry \cite{Azzurli:2017kxo,Bobev:2017uzs}. For the universal twist, the dual supergravity background corresponds to the Euclidean AdS-Taub-Bolt solution in the 4d $\mN=2$ minimal gauged supergravity. This background has already been studied extensively in the context of holography \cite{Toldo:2017qsh,Bobev:2020zov}, and notably, its regularized on-shell action has been successfully matched with the $\mM_{\mg,p}$ partition function of ABJM theory in the large $N$ limit. Hence we refer the readers to \cite{Toldo:2017qsh,Bobev:2020zov} as well as earlier works such as \cite{Martelli:2012sz,Chamblin:1998pz,Hawking:1998ct,Alonso-Alberca:2000zeh,Taub:1950ez,Newman:1963yy,Gibbons:1979xm} for a comprehensive analysis of the background concerning its global structure, the distinctions between NUTs and bolts, supersymmetry, moduli spaces, and constraints for uplift to 11d supergravity. 

\medskip

In the following subsections we would rather focus on presenting the well-known Euclidean AdS-Taub-Bolt background in a slightly different convention from those in \cite{Toldo:2017qsh,Bobev:2020zov}. This adjustment will facilitate the holographic matching discussed in section \ref{sec:holo} not only between the regularized on-shell action and the $\mM_{\mg,p}$ partition function but also between the 4d supergravity background in the asymptotic regime and the 3d new minimal supergravity background reviewed in subsection \ref{sec:Mgp:bg}. While this latter comparison was initiated in \cite{Toldo:2017qsh}, the flat connection in the 4d graviphoton was not specified there, which is essential to determine the U(1)$_R$ holonomy $\nu_R$ unambiguously on the field theory side. We address this issue by specifying a proper choice of the graviphoton flat connection, which is the main goal of this section. This choice ensures that the U(1)$_R$ holonomy in the boundary field theory takes the specific values $\nu_R\in\{\pm\fft12\}$, which will be examined in the next section \ref{sec:holo}. 

\subsection{Euclidean AdS-Taub-Bolt geometry}\label{sec:sugra:bolt}
We mainly follow the conventions of \cite{Toldo:2017qsh,Bobev:2020zov} to describe the Euclidean AdS-Taub-Bolt solution in the 4d $\mN=2$ minimal gauged supergravity, but restoring the AdS radius $L$ and reparametrizing the Euclidean time coordinate, radial coordinate, and squashing parameter as
\begin{align}
	(\tau,r,s)\to L(\tau,r,s)\,,\label{convention}
\end{align}
to ensure that $(\tau,r,s)$ remain dimensionless. 

\subsubsection{Local solution}\label{sec:sugra:bolt:local}
The 4d Euclidean $\mN=2$ minimal gauged supergravity has the bosonic action
\begin{align}
	S_\text{bulk}=-\fft{1}{16\pi G_N}\int d^4x\,\sqrt{g}\Big[R-2\Lambda-F_{\mu\nu}F^{\mu\nu}\Big]\qquad\bigg(\Lambda=-\fft{3}{L^2}\bigg)\,,\label{S:bulk}
\end{align}
and the corresponding equations of motion read
\begin{subequations}
	\begin{align}
		R_{\mu\nu}-\fft12g_{\mu\nu}(R-2\Lambda)&=2F_\mu{}^\rho F_{\nu\rho}-\fft12g_{\mu\nu}F_{\rho\sigma}F^{\rho\sigma}\,,\\
		\nabla_\mu F^{\mu\nu}&=0\,.
	\end{align}\label{eom}%
\end{subequations}

The Euclidean AdS-Taub-Bolt/NUT solution to the equations of motion (\ref{eom}) is described locally as
\begin{subequations}
	\begin{align}
		ds^2&=L^2\bigg[\lambda(r)(d\tau-2sf_\kappa(\theta)d\phi)^2+\fft{dr^2}{\lambda(r)}+(r^2-s^2)ds^2_{\Sigma_{\mg}}\bigg]\,,\label{Taub:metric}\\
		A&=\fft{P(r^2+s^2)-2sQr}{2sL(r^2-s^2)}(d\tau-2sf_\kappa(\theta)d\phi)+\alpha d\tau\,,\label{Taub:graviphoton}
	\end{align}\label{Taub}%
\end{subequations}
where we have used the expressions (\ref{f:Sigma}) and the radial function
\begin{align}
	\lambda(r)=\fft{(r^2-s^2)^2+(\kappa-4s^2)(r^2+s^2)-2\fft{M}{L}r+\fft{P^2-Q^2}{L^4}}{r^2-s^2}\,.\label{lambda}
\end{align}
Notably, we have included a flat connection $\alpha d\tau$ in the graviphoton (\ref{Taub:graviphoton}) that will be specified below through a regularity condition.

\subsubsection{Global properties}\label{sec:sugra:bolt:global}
The Euclidean AdS-Taub-Bolt solution is locally described by (\ref{Taub}) and is further characterized by a fixed 2-dimensional submanifold (``bolt'') for the Killing vector $\partial_\tau$ at a radius $r=r_b>s$ such that
\begin{align}
	\lambda(r_b)=0\,,\qquad\lambda'(r_b)>0\,,\qquad \lambda(r>r_b)>0\,.\label{rb}
\end{align}
Near the bolt $r=r_b$, the metric (\ref{Taub:metric}) takes the form
\begin{align}
	ds^2\Big|_{r\to r_b}\approx L^2\bigg[\fft{4}{\lambda'(r_b)}\bigg[dx^2+x^2\fft{\lambda'(r_b)^2}{4}(d\tau-2sf_\kappa d\phi)^2\bigg]+(r_b^2-s^2)ds^2_{\Sigma_{\mg}}\bigg]\,,\label{Taub:metric:bolt}
\end{align}
where $x\equiv\sqrt{r-r_b}$. The absence of a conical singularity then relates the Euclidean time period to the first derivative of the radial function at the bolt as
\begin{equation}
	\tau\sim\tau+\beta_\text{grav}\quad\to\quad\lambda'(r_b)=\fft{4\pi}{\beta_\text{grav}}\,.\label{lambdaprime}
\end{equation}

On the other hand, we also require that the asymptotic boundary $ds^2_{3,\text{grav}}$ extracted from the metric (\ref{Taub:metric}) in the $r\to\infty$ limit, namely
\begin{equation}
	ds^2\Big|_{r\to\infty}\approx L^2\bigg[\fft{dr^2}{r^2}+r^2\underbrace{\Big[(d\tau-2sf_\kappa(\theta)d\phi)^2+ds^2_{\Sigma_{\mg}}\Big]}_{=ds_{3,\text{grav}}^2}\bigg]\,,\label{Taub:metric:asymp}
\end{equation}
corresponds to the $\mM_{\mg,p}$ background reviewed in subsection \ref{sec:Mgp:bg} for holographic analysis. This boundary condition determines the Euclidean time period as \cite{Martelli:2012sz,Toldo:2017qsh,Bobev:2020zov}
\begin{equation}
	\beta_\text{grav}=\fft{8\pi\eta s}{p}\label{beta:grav}
\end{equation}
in terms of $\eta$ introduced in (\ref{f:Sigma}). Note that the Euclidean time period $\beta_\text{grav}$ must satisfy both (\ref{lambdaprime}) and (\ref{beta:grav}) for the metric (\ref{Taub:metric}) to represent a regular Euclidean AdS-Taub-Bolt background. As a consequence, we obtain
\begin{equation}
	\lambda'(r_b)=\fft{p}{2\eta s}>0\,,\label{lambdaprime+betagrav}
\end{equation}
which requires that the parameters $p$ and $s$ have the same sign, given that $\eta$ is positive as defined in (\ref{f:Sigma}).

\medskip

The last regularity condition pertains to the graviphoton at the bolt. Specifically, imposing the regularity constraint $A_\mu A^\mu|_{r=r_b}<\infty$ fixes the flat connection $\alpha d\tau$ introduced in (\ref{Taub:graviphoton}) by
\begin{equation}
	\alpha=-\fft{P(r_b^2+s^2)-2sQr_b}{2sL(r_b^2-s^2)}\,.\label{alpha}
\end{equation}
Here we follow the prescription outlined in \cite{Cabo-Bizet:2018ehj,BenettiGenolini:2023rkq,BenettiGenolini:2023ucp} to establish the flat connection. For additional context, see \cite{Martelli:2012sz}, which discusses the necessity of the flat connection based on the requirement for a well-defined 1-form on a given background equipped with appropriate coordinate patches.

\subsubsection{Supersymmetry}\label{sec:sugra:bolt:susy}
The Euclidean AdS-Taub-Bolt geometry described above can be made to preserve certain amount of supersymmetry by solving the Killing spinor equation \cite{Romans:1991nq}
\begin{equation}
	\bigg(\partial_\mu+\fft14\omega_\mu^{ab}\gamma_{ab}+\fft{1}{2L}\gamma_\mu-\fft{\ri}{L}A_\mu+\fft{\ri}{4}F_{\nu\rho}\gamma^{\nu\rho}\gamma_\mu\bigg)\epsilon=0\,.\label{KSE}
\end{equation}
We are particularly interested in the $\fft14$-BPS solution that preserves two real supercharges, as we aim to evaluate the holographic dual partition function on the supersymmetric $\mM_{\mg,p}$ background keeping the same number of real supercharges. For the $\fft14$-BPS background, the integrability conditions arising from the Killing spinor equation (\ref{KSE}) yield the following two constraints \cite{Martelli:2012sz,Toldo:2017qsh,Bobev:2020zov}
\begin{equation}
	P=-\fft{L^2}{2}(4s^2-\kappa)\qquad\&\qquad M=\fft{2sQ}{L}\,.\label{integrability}
\end{equation}
See Appendix \ref{app:conventions} for conventions used for the Killing spinor equations (\ref{KSE}).

\medskip

Under the integrability conditions (\ref{integrability}), the Euclidean AdS-Taub-Bolt geometry can be categorized into two distinct classes denoted by Bolt$_\pm$. To begin with, note that the radial function (\ref{lambda}) is factored under the integrability conditions (\ref{integrability}) explicitly as
\begin{equation}
	\lambda(r)\Big|_\text{(\ref{integrability})}=\fft{\overbrace{(r^2-2sr+\fft{\kappa}{2}-s^2-\fft{Q}{L^2})}^{=(r-r_+^{(+)})(r-r_+^{(-)})}\overbrace{(r^2+2sr+\fft{\kappa}{2}-s^2+\fft{Q}{L^2})}^{=(r-r_-^{(+)})(r-r_-^{(-)})}}{r^2-s^2}\,,\label{lambda:fact}
\end{equation}
where
\begin{equation}
	r_+^{(\pm)}=s\pm\sqrt{-\fft{\kappa}{2}+2s^2+\fft{Q^2}{L^2}}\,,\qquad r_-^{(\pm)}=-s\pm\sqrt{-\fft{\kappa}{2}+2s^2-\fft{Q^2}{L^2}}\,.
\end{equation}
Bolt$_\pm$ solutions are then determined by the location of the bolt $r=r_b$ as
\begin{equation}
	r_b=r_\pm^{(+)}>r_\mp^{(+)}\,,
\end{equation}
where the inequality ensures that $r_\pm^{(+)}$ is the largest positive root of the radial function.

\medskip

The key expressions governing the Euclidean AdS-Taub-Bolt geometry can be written more explicitly for the $\fft14$-BPS Bolt$_\pm$ solutions defined above as follows.
\begin{itemize}
	\item The electric charge parameter $Q$ for the Bolt$_\pm$ solutions is given by
	\begin{subequations}
	\begin{align}
		Q&=Q_+=L^2\fft{p^2\pm\sqrt{(p-16\eta s^2)^2\big((p+16\eta s^2)^2-128\kappa\eta^2 s^2\big)}}{128\eta^2s^2}\,,\\
		Q&=Q_-=-L^2\fft{p^2\pm\sqrt{(p+16\eta s^2)^2\big((p-16\eta s^2)^2-128\kappa\eta^2 s^2\big)}}{128s^2}\,,
	\end{align}\label{Q:boltpm}%
	\end{subequations}
	respectively. To derive these expressions, we have used (\ref{lambdaprime+betagrav}) written explicitly as
	\begin{equation}
		\lambda'(r_b)=\fft{2((r_\pm^{(+)})^2\pm2sr_\pm^{(+)}+\fft{\kappa}{2}-s^2\pm \fft{Q}{L^2})}{r_\pm^{(+)}\pm s}=\fft{p}{2\eta s}\label{lambdaprime+betagrav:boltpm}
	\end{equation}
	derived from the factorization (\ref{lambda:fact}).
	
	\item Substituting the electric charge parameters $Q_\pm$ back into the bolt locations $r_b=r_\pm^{(+)}$, we obtain
	\begin{subequations}
	\begin{align}
		r_b&=r_+^{(+)}=s+\bigg|\fft{\sqrt{(p-16\eta s^2)^2}\pm\sqrt{(p+16\eta s^2)^2-128\kappa\eta^2 s^2}}{16\eta^2s}\bigg|\,,\\
		r_b&=r_-^{(+)}=-s+\bigg|\fft{\sqrt{(p+16\eta s^2)^2}\pm\sqrt{(p-16\eta s^2)^2-128\kappa\eta^2 s^2}}{16\eta^2s}\bigg|\,.
	\end{align}
	\end{subequations}
	The $\pm$ signs are in the same order as in (\ref{Q:boltpm}), which are independent of the choice of the Bolt$_\pm$ backgrounds.
	
	\item The $\alpha$ term in (\ref{alpha}), which governs the graviphoton flat connection $\alpha d\tau$, can be simplified significantly by employing the expressions (\ref{Q:boltpm}) and (\ref{lambdaprime+betagrav:boltpm}) as
	\begin{equation}
		\alpha\Big|_\text{(\ref{integrability})}=-L\bigg(\fft{\kappa}{4s}\mp\fft{p}{8\eta s}\bigg)
	\end{equation}
	for the Bolt$_\pm$ solutions respectively.
\end{itemize}
The necessary condition for a regular Bolt$_\pm$ solution can be inferred from the above expressions as 
\begin{align}
	(p\pm 16\eta s^2)^2-128\kappa\eta^2s^2\geq0\,,
\end{align}
which we will assume from here on. Other necessary conditions presented in \cite{Toldo:2017qsh} have already been imposed above. 

\medskip

Finally, the solution to the Killing spinor equation (\ref{KSE}) on the Bolt$_\pm$ backgrounds can be found explicitly as 
\begin{equation}
	\epsilon=e^{\pm\fft{\ri p}{8\eta s}\tau}\begin{pmatrix}
		0 \\ \sqrt{\fft{(r-r_+^{(+)})(r-r_+^{(-)})}{r-s}}\chi \\ 0 \\ \ri\sqrt{\fft{(r-r_-^{(+)})(r-r_-^{(-)})}{r+s}}\chi
	\end{pmatrix}\label{epsilon}
\end{equation}
in terms of a constant complex parameter $\chi$ with two real degrees of freedom. We derived the above Killing spinor in the static vielbein
\begin{equation}
	\begin{alignedat}{2}
		e^1&=L\sqrt{r^2-s^2}d\theta\,,&\qquad e^2&=L\sqrt{r^2-s^2}f_\kappa'(\theta)d\phi\,,\\ e^3&=L\lambda(r)^{1/2}(d\tau-2sf_\kappa(\theta)d\phi)\,,&\qquad e^4&=L\lambda(r)^{-1/2}dr\,,
	\end{alignedat}
\end{equation}
and followed the gamma matrix convention summarized in Appendix \ref{app:conventions}. Note that differences in conventions result in a slightly different Killing spinor compared to those of \cite{Martelli:2012sz,Toldo:2017qsh}. In particular, the Killing spinor (\ref{epsilon}) exhibits a non-trivial $\tau$-dependence, which will be compared with the $\psi$-dependence of the 3d Killing spinor through holography in section \ref{sec:holo}.

\subsection{Regularized on-shell action}\label{sec:sugra:onshell}
To evaluate the regularized on-shell action of the EAdS-Taub-Bolt solution, we first introduce the Gibbons-Hawking-York boundary term and the counterterms as 
\begin{subequations}
\begin{align}
	S_\text{GHY}&=-\fft{1}{8\pi G_N}\int d^3x\,\sqrt{h}K\,,\\
	S_\text{ct}&=\fft{1}{8\pi G_N}\int d^3x\,\sqrt{h}\bigg[\fft{2}{L}+\fft{L}{2}R^{(3)}\bigg]\,,
\end{align}\label{S:bdry}%
\end{subequations}
where 3d metric $h_{ij}$ is the pull back of the Euclidean AdS-Taub-Bolt background (\ref{Taub:metric}) at the asymptotic boundary $r=r_\infty(\to\infty)$, $R^{(3)}$ is the associated Ricci scalar, and $K$ is the extrinsic curvature defined as
\begin{equation}
	K_{\mu\nu}=\nabla_\mu n_\nu-n_\mu n^\rho\nabla_\rho n_\nu\quad\to\quad K=K_\mu{}^\mu
\end{equation}
with $n^\mu$ being the outward unit normal vector to the asymptotic boundary. 

\medskip

The regularized on-shell action of the EAdS-Taub-Bolt solution can be evaluated by substituting the background (\ref{Taub}) into the actions (\ref{S:bulk}) and (\ref{S:bdry}) as 
\begin{align}
	&\Big[S_\text{bulk}+S_\text{GHY}+S_\text{ct}\Big]_\text{(\ref{Taub})}\nn\\
	&=\bigg[\fft{2M}{L}-2r_b^3+6r_bs^2+\fft{(P-Q)^2}{(r_b-s)^2}\fft{r_b}{L^4}+\fft{(P+Q)^2}{(r_b+s)^2}\fft{r_b}{L^4}\bigg]\fft{\eta sL^2\text{vol}[\Sigma_{\mg}]}{2p G_N}\,,\label{S:total}
\end{align}
where we have also used the global properties presented in subsection \ref{sec:sugra:bolt:global}. For the $\fft14$-BPS Bolt$_\pm$ backgrounds of interest, the regularized on-shell action (\ref{S:total}) simplifies significantly to \cite{Toldo:2017qsh,Bobev:2020zov}
\begin{align}
	I_{2\partial,\text{Bolt}_\pm}\equiv\Big[S_\text{bulk}+S_\text{GHY}+S_\text{ct}\Big]_\text{(\ref{Taub}),\,Bolt$_\pm$}=\fft{\pi L^2}{8G_N}\Big[4(1-\mg)\mp p\Big]\,,\label{I:2der}
\end{align}
where various supersymmetric properties spelled out in subsection \ref{sec:sugra:bolt:susy} have been employed. 

\medskip

The 4-derivative corrections to the regularized on-shell action of the Bolt$_\pm$ solution (\ref{I:2der}) have been analyzed in \cite{Bobev:2020zov} based on the structure of supersymmetric 4-derivative corrections to the 4d $\mN=2$ minimal gauged supergravity \cite{Bobev:2020egg,Bobev:2021oku}. The results can be summarized alongside the 2-derivative on-shell action (\ref{I:2der}) as \cite{Bobev:2020zov}
\begin{align}
	I_{2\partial,\text{Bolt}_\pm}+I_{4\partial,\text{Bolt}_\pm}=\bigg[\fft{\pi L^2}{2G_N}+32\pi^2(c_2-c_1)\bigg]\fft{4(1-\mg)\mp p}{4}+64\pi^2c_1(1-\mg)\,,\label{I:2der+4der}
\end{align}
where the two constant coefficients $c_{1,2}$ are associated with the supersymmetric completions of the Weyl square term and the Gauss-Bonnet term governing the 4-derivative bosonic actions. We refer the readers to \cite{Bobev:2020egg,Bobev:2021oku,Bobev:2020zov,Genolini:2021urf} for more detailed explanation concerning the supersymmetric 4-derivative corrections and their on-shell evaluation. In the next section \ref{sec:holo}, we will compare the known result (\ref{I:2der+4der}) with the field theory dual quantity examined in the previous section \ref{sec:ABJM}.

\section{Comments on holography}\label{sec:holo}
In this section we compare the field theory results from section \ref{sec:ABJM} and the supergravity results from section \ref{sec:sugra} through holography. Subsection \ref{sec:holo:bg} provides a holographic comparison of the backgrounds, highlighting the importance of the graviphoton flat connection specified in the previous section \ref{sec:sugra}. We then discuss the holographic matching between the partition functions in subsection \ref{sec:holo:ptnfct}, refining the previous work \cite{Toldo:2017qsh,Bobev:2020zov}.

\subsection{Backgrounds}\label{sec:holo:bg}
In the context of holographic duality, the $\fft14$-BPS Bolt$_\pm$ backgrounds reviewed in subsection \ref{sec:sugra:bolt} are expected to correspond to the supersymmetric background $\mM_{\mg,p}$ examined in subsection \ref{sec:Mgp:bg} in the asymptotic region. To demonstrate this explicitly, we first present the asymptotic behavior of the Bolt$_\pm$ backgrounds as
\begin{subequations}
	\begin{align}
		ds^2\Big|_{\text{Bolt}_\pm,\,r\to\infty}&=L^2\bigg[\fft{dr^2}{r^2}+r^2\underbrace{\Big[(d\tau-2sf_\kappa(\theta)d\phi)^2+ds^2_{\Sigma_{\mg}}\Big]}_{=ds^2_{3,\text{grav}}}\bigg]\,,\label{metric:asymp}\\
		A\Big|_{\text{Bolt}_\pm,\,r\to\infty}&=\fft{P}{2sL}(d\tau-2sf_\kappa(\theta)d\phi)-\fft{P(r_b^2+s^2)-2sQr_b}{2sL(r_b^2-s^2)} d\tau\nn\\
		&=L\bigg(\fft{\kappa}{4s}-s\bigg)\bigg(d\tau-2sf_\kappa(\theta)d\phi\bigg)- L\bigg(\fft{\kappa}{4s}\mp\fft{p}{8\eta s}\bigg)d\tau\,,\label{graviphoton:asymp}
	\end{align}\label{bolt:asymp}%
\end{subequations}
where in the last line we have used the properties of the Bolt$_\pm$ solutions outlined in subsection \ref{sec:sugra:bolt:susy}. Then, by identifying the 4d and 3d coordinates \& parameters as
\begin{equation}
	\tau=-\fft{4\eta s}{p}\psi\qquad\&\qquad \fft{4\eta s^2}{p^2}=\beta_\text{3d}^2\,,\label{4d:3d}
\end{equation}
we confirm that the asymptotic behaviors of the Bolt$_\pm$ backgrounds (\ref{bolt:asymp}) align with the 3d supersymmetric $\mM_{\mg,p}$ background presented in (\ref{Seifert:metric}) and (\ref{Seifert:AHV}) as
\begin{subequations}
\begin{align}
	ds_{3,\text{grav}}^2&=4\eta ds_3^2\,,\label{bg:matching:metric}\\
	A\Big|_{\text{Bolt}_\pm,\,r\to\infty}&=LA^{(R)}\Big|_{\ell=\mp\fft12\psi}\,.\label{bg:matching:A}
\end{align}\label{bg:matching}%
\end{subequations}
The following comments pertain to the matching of these backgrounds.
\begin{itemize}
	\item 3d backgrounds (\ref{Seifert:metric}) and (\ref{Seifert:AHV}) are introduced without the explicit length scale and therefore appropriate powers of the AdS radius $L$ are extracted from the 4d backgrounds for matching.
	
	\item In (\ref{bg:matching:metric}), the factor of $4\eta$ difference arises from the choice of convention, where the volume of the 2d Riemann surface within the 3d background (\ref{Seifert:metric}) is given by $\pi$ \cite{Closset:2017zgf,Closset:2018ghr,Closset:2019hyt}, while we have used $\vol[\Sigma_{\mg}]=4\pi\eta$.
	
	\item In (\ref{bg:matching:A}), the matching \emph{requires} the U(1)$_R$ holonomy $\nu_R$, which governs the flat connection in (\ref{Seifert:AHV:A}) through $\ell=-\nu_R\psi$, to assume specific values $\nu_R=\pm\fft12$ for the Bolt$_\pm$ solutions respectively. This means that, although the U(1)$_R$ holonomy $\nu_R$ can generally take integer or half-integer values on the field theory side, it must specifically be set as $\nu_R=\pm\fft12$ to achieve precise holographic matching with the dual supersymmetric Euclidean AdS-Taub-Bolt$_\pm$ backgrounds. This finding generalizes the work of \cite{BenettiGenolini:2023ucp}, which identifies a similar feature in the holographic duality between the $S^1\times\Sigma_{\mg}$ TTI and the dual magnetically charged Reissner-Nordstr\"om AdS black hole. See also \cite{BenettiGenolini:2023rkq,Cabo-Bizet:2018ehj} for similar observations in different contexts of AdS/CFT.
\end{itemize}

The Killing spinor equation (\ref{KSE}) for the 4d $\mN=2$ minimal gauged supergravity can also be reduced in the asymptotic regime ($r\to\infty$) and compared with the 3d generalized Killing spinor equations (\ref{3d:KSE}) for the new minimal supergravity in three dimensions. For details on the reduction of the Killing spinor equation, we refer readers to \cite{Martelli:2012sz,Toldo:2017qsh}. A key difference in this paper is that the Killing spinor (\ref{epsilon}) constructed in the static vielbein has a non-trivial dependence on the Euclidean time coordinate $\tau$, which is related to the Seifert fiber coordinate $\psi$ through (\ref{4d:3d}). Specifically, we have
\begin{align}
	\epsilon=\begin{pmatrix}
		\sqrt{\fft{(r-r_+^{(+)})(r-r_+^{(-)})}{r-s}}  \\ \ri\sqrt{\fft{(r-r_-^{(+)})(r-r_-^{(-)})}{r+s}}
	\end{pmatrix}\otimes\underbrace{e^{\pm\fft{\ri p}{8\eta s}\tau}\begin{pmatrix}
		0 \\ \chi
		\end{pmatrix}}_{=\zeta}\qquad\xrightarrow{\text{(\ref{4d:3d})}}\qquad \zeta=e^{\mp\fft{\ri}{2}\psi}\begin{pmatrix}
		0 \\ \chi
	\end{pmatrix}\,.
\end{align}
This implies that the dual supergravity Bolt$_\pm$ backgrounds require the fermions in the boundary field theory to be anti-periodic along the Seifert fiber with the U(1) coordinate $\psi\sim\psi+2\pi$. This requirement is directly related to the specification of the flat connection in (\ref{bg:matching:A}), see \cite{BenettiGenolini:2023ucp,BenettiGenolini:2023rkq,Cabo-Bizet:2018ehj} for related discussions.

\subsection{Partition functions}\label{sec:holo:ptnfct}
To compare the $\mM_{\mg,p}$ partition function (\ref{ABJM:logZ:result}) with its holographic dual in the 4d $\mN=2$ minimal gauged supergravity reviewed in section \ref{sec:sugra}, we first implement the universal twist by turning off all holonomies and magnetic fluxes associated with the flavor symmetries except for the superconformal $R$ symmetry \cite{Azzurli:2017kxo,Bobev:2017uzs}. This corresponds to the choice
\begin{equation}
	\Delta_a^*=\fft12\nu_R\qquad\&\qquad \mn_a^*=\fft{1-\mg}{2}\label{universal}
\end{equation}
according to the reparametrization (\ref{ABJM:repara}). Thanks to the dual supergravity background analysis in the previous subsection \ref{sec:holo:bg} that specifies the U(1) holonomy as $\nu_R\in\{\pm\fft12\}$ for the Bolt$_\pm$ backgrounds respectively, the universal twist (\ref{universal}) leads to
\begin{equation}
	[\Delta_a^*]=\begin{cases}
		\fft14 & (\nu_R=\fft12,~\text{Bolt}_+) \\
		\fft34 & (\nu_R=-\fft12,~\text{Bolt}_-) 
	\end{cases}\,.\label{universal:shift}
\end{equation}
Importantly, the fact that (\ref{universal}) and (\ref{universal:shift}) describe the universal twist dual to the 4d minimal gauged supergravity is no longer merely conjectural as in \cite{Toldo:2017qsh}, and is determined unambiguously based on the reparametrization (\ref{ABJM:repara}) and the U(1)$_R$ holonomy specified by the graviphoton flat connection. 

\medskip

Substituting (\ref{universal}) and (\ref{universal:shift}) into (\ref{ABJM:logZ:result}), we obtain the $\mM_{\mg,p}$ partition function of ABJM theory at the universal twist as
\begin{align}
	\Re\log Z^\text{ABJM}_{\mM_{\mg,p}}(\bDelta^*,\bmn^*)&=\pm p\bigg[\fft{\pi\sqrt{2k}}{12}\bigg(N-\fft{k}{24}+\fft{2}{3k}\bigg)^{3/2}+\hat{g}_0(k,\bDelta^*)\bigg]\nn\\
	&\quad-\fft{\pi(1-\mg)\sqrt{2k}}{3}\bigg[\bigg(N-\fft{k}{24}+\fft{2}{3k}\bigg)^{3/2}-\fft3k\bigg(N-\fft{k}{24}+\fft{2}{3k}\bigg)^{1/2}\bigg]\nn\\
	&\quad-\fft{1-\mg}{2}\log\bigg(N-\fft{k}{24}+\fft{2}{3k}\bigg)+\hat f_0(k,\bDelta^*,\bmn^*)+\mO(e^{-\sqrt{N}})\label{ABJM:logZ:result:universal}
\end{align}
for the cases dual to the Bolt$_\pm$ backgrounds respectively. To derive the result dual to the Bolt$_-$ solution, in particular, we have used the symmetry discussed around (\ref{ABJM:Z:symmetry}). 

\medskip

According to the AdS/CFT correspondence, the above $\mM_{\mg,p}$ partition function of ABJM theory (\ref{ABJM:logZ:result:universal}) is supposed to capture the Euclidean path integral of M-theory over the 11d background constructed by uplifting the Euclidean AdS-Taub-Bolt$_\pm$ geometry through the consistent truncation on the $S^7/\mathbb{Z}_k$ internal manifold \cite{deWit:1986oxb,Gauntlett:2007ma}.\footnote{In this paper we implicitly assume that the Euclidean AdS-Taub-Bolt$_\pm$ geometry provides the dominant contribution to the M-theory path integral. For details concerning the non-trivial moduli space of AdS-Taub solutions and the necessary conditions for the Euclidean AdS-Taub-Bolt$_\pm$ geometry to serve as a dominant saddle, see \cite{Martelli:2012sz,Toldo:2017qsh} for example.} In the semi-classical limit where the 4d Newton constant $G_N$ is much smaller than the square of the AdS radius $L^2$, this M-theory path integral can be approximated by the regularized on-shell action of the 4d $\mN=2$ minimal gauged supergravity as \cite{Drukker:2010nc}
\begin{equation}
	\Re\log Z^\text{M-theory}\Big|_{\text{Bolt}_\pm\times_\text{f} S^7/\mathbb{Z}_k}=-I_{2\partial,\text{Bolt}_\pm}-I_{4\partial,\text{Bolt}_\pm}+\mO\Big(\log\fft{L^2}{G_N}\Big)\,,\label{M:logZ}
\end{equation}
where we have also included the 4-derivative corrections briefly discussed in subsection \ref{sec:sugra:onshell} and focused on the real part that is free from the branch choice ambiguity. In (\ref{M:logZ}) the subscript ``f'' indicates that the 11d background involves a non-trivial fibration. It is then straightforward to check that the $\mM_{\mg,p}$ partition function of ABJM theory (\ref{ABJM:logZ:result:universal}) and the dual M-theory path integral (\ref{M:logZ}) are consistent with each other in the first two leading terms of orders $N^{3/2}$ and $N^{1/2}$ under the AdS/CFT dictionary \cite{Bobev:2021oku}\footnote{The parameter $a$ is not fully determined relying solely on the holographic matching \cite{Bobev:2021oku} but its value was recently proposed based on the all order closed-form expression for the ABJM TTI in \cite{Bobev:2022eus}.}
\begin{equation}
	\fft{L^2}{2G_N}=\fft{\sqrt{2k}}{3}N^{3/2}+aN^{1/2}\,,\quad c_1=-\fft{1}{\sqrt{2k}}\fft{1}{32\pi}N^{1/2}\,,\quad c_2=-\bigg(a+\fft{k^2+8}{24\sqrt{2k}}\bigg)\fft{1}{32\pi}N^{1/2}\,,\label{dictionary}
\end{equation}
where we have used the regularized on-shell actions presented in (\ref{I:2der+4der}). The above matching between (\ref{ABJM:logZ:result:universal}) and (\ref{M:logZ}) improves the previous holographic comparison restricted to the $N^{3/2}$ leading order \cite{Toldo:2017qsh} by incorporating the first subleading corrections of order $N^{1/2}$, supporting the supergravity analysis for the first subleading corrections \cite{Bobev:2020zov} directly from the field theory perspective. For holographic comparison at the $N^{3/2}$ leading order beyond the universal/minimal case, we refer the readers to \cite{BenettiGenolini:2024xeo,BenettiGenolini:2024hyd} based on equivariant localization circumventing the absence of explicit solutions \cite{BenettiGenolini:2023kxp,BenettiGenolini:2023ndb,Martelli:2023oqk,Colombo:2023fhu}.

\medskip

The next-to-next-to-leading order (NNLO) correction in the Euclidean M-theory path integral (\ref{M:logZ}) is expected to be logarithmic, behaving as $\sim\log\relax(L^2/G_N)$. This logarithmic correction has not yet been derived from first principles, primarily due to the subtlety of zero-mode counting in a generic asymptotically Euclidean AdS$_4$ background, which includes the Euclidean AdS-Taub-Bolt solution of our interest, within the 1-loop analysis for the gravitational path integral. For similar 1-loop analyses in different backgrounds, see \cite{Bhattacharyya:2012ye,Liu:2017vbl,Benini:2019dyp,PandoZayas:2020iqr}. It has recently been claimed, however, that the logarithmic correction is characterized by a pure number coefficient governed by the Euler characteristic of a given 4d background $\mM_4$ as \cite{Bobev:2023dwx}
\begin{align}
	\Re\log Z^\text{M-theory}\Big|_{\mM_4\times_\text{f} S^7/\mathbb{Z}_k}&=-I_{2\partial,\mM_4}-I_{4\partial,\mM_4}-\fft16\chi(\mM_4)\log\fft{L^2}{G_N}+\mO(1)\,,\label{log:proposal}\\[0.5em]
	\chi(\mM_4)&=\fft{1}{32\pi^2}\int_{\mM_4} d^4x\,\sqrt{g}\,(R_{\mu\nu\rho\sigma}R^{\mu\nu\rho\sigma}-4R_{\mu\nu}R^{\mu\nu}+R^2)\,.\nn
\end{align}
See also \cite{Hristov:2021zai} where the proportionality between the coefficient of the logarithmic term and the Euler characteristic is first analyzed employing the Atiyah-Singer index theorem. For the Euclidean AdS-Taub-Bolt solution of interest with the Euler characteristic $\chi=2(1-\mg)$,\footnote{There is a subtle issue in evaluating the Euler characteristic for the Euclidean AdS-Taub-Bolt geometry due to the non-trivial boundary contribution \cite{Hristov:2021zai}. The successful matching of $\log N$ order corrections across (\ref{ABJM:logZ:result:universal}) and (\ref{M:logZ:log}) suggests that this subtlety does not undermine the overall picture. Addressing why the boundary contribution to the Euler characteristic of the Euclidean AdS-Taub-Bolt geometry can be disregarded will be left for future work.} the above formulation translates to
\begin{equation}
	\Re\log Z^\text{M-theory}\Big|_{\text{Bolt}_\pm\times_\text{f} S^7/\mathbb{Z}_k}=-I_{2\partial,\text{Bolt}_\pm}-I_{4\partial,\text{Bolt}_\pm}-\fft{1-\mg}{2}\log N+\mO(1)\,,\label{M:logZ:log}
\end{equation}
where we have also employed the AdS/CFT dictionary (\ref{dictionary}). The $\log N$ term in the $\mM_{\mg,p}$ partition function (\ref{ABJM:logZ:result:universal}) then perfectly matches the NNLO term in (\ref{M:logZ:log}) as determined by the proposal (\ref{log:proposal}). In this context, our result for the $\mM_{\mg,p}$ partition function can also support the analysis of \cite{Bobev:2023dwx} regarding the logarithmic correction in the Euclidean M-theory path integral.

\medskip

To be precise, the above holographic comparison is valid only if the magnetic flux associated with the background graviphoton satisfies a certain quantization condition necessary for uplifting the Euclidean AdS-Taub-Bolt geometry to 11d supergravity \cite{Martelli:2012sz,Toldo:2017qsh}. We implicitly assume this necessary condition and direct the readers to \cite{Martelli:2012sz,Toldo:2017qsh} for further details. Note that the flat connection governed by the $\ell$-function of the form $\ell=-\nu_R\psi$ in (\ref{Seifert:AHV:A}) affects only the holonomy along the Seifert fiber, rather than the magnetic flux on the Riemann surface, allowing us to rely on the same quantization condition.

\section{Discussion}\label{sec:discussion}
In this paper we study the supersymmetric partition function of the $\text{U}(N)_k\times \text{U}(N)_{-k}$ ABJM theory on the Seifert manifold $\mM_{\mg,p}$ describing a U(1) bundle over a Riemann surface of genus $\mg$ with first Chern number $p\in\mathbb{Z}$ and investigate its holographic implications. The ABJM $\mM_{\mg,p}$ partition function is evaluated to all orders in the $1/N$-perturbative expansion by applying results from the numerical analysis of the ABJM TTI \cite{Bobev:2022jte,Bobev:2022eus} in the context of the Bethe formulation \cite{Closset:2016arn,Closset:2017zgf,Closset:2018ghr,Closset:2019hyt}. On the dual supergravity side, we revisit the Euclidean AdS-Taub-Bolt solution in 4d $\mN=2$ minimal gauged supergravity, specifying the flat connection in the background graviphoton. This supergravity background and the corresponding regularized on-shell action have been successfully identified with the 3d supersymmetric background and the $\mM_{\mg,p}$ partition function of ABJM theory respectively. In due process, the graviphoton flat connection determines the U(1)$_R$ holonomy in the boundary field theory unambiguously, which is crucial for achieving precise holographic matching. In our holographic analysis, we clarify how the previous comparison limited to the $N^{3/2}$ leading order \cite{Toldo:2017qsh} is improved to incorporate the first subleading corrections of order $N^{1/2}$, determined by the supergravity analysis \cite{Bobev:2020egg,Bobev:2020zov,Bobev:2021oku}, and to support the analysis regarding the $\log N$ correction in the Euclidean M-theory path integral \cite{Bobev:2023dwx,Hristov:2021zai}.

\medskip

The list of selected open questions for future research is as follows.

\medskip

The next logical step is to generalize our work to other $\mN\geq 2$ holographic SCFTs arising from the $N$ stack of M2-branes located at the tip of the cone over 7-dimensional Sasaki-Einstein manifolds including $S^7/\mathbb{Z}_k$ (with a fixed point, which is distinguished from the smooth $S^7/\mathbb{Z}_k$ associated with ABJM theory), $N^{0,1,0}$, $Q^{1,1,1}$, $V^{5,2}$, see \cite{Hosseini:2016ume,Bobev:2023lkx,Bobev:2024mqw} and references therein. While closed-form expressions for their TTI and effective twisted superpotentials are already available in the literature \cite{Bobev:2023lkx,Bobev:2024mqw}, the latter are primarily provided for the universal twist in many cases. Once this gap is addressed, the subsequent steps will closely mirror the analysis presented in this paper. 

Another natural question concerns the analytic derivation of an all order closed-form expression for the ABJM $\mM_{\mg,p}$ partition function. A similar question has been discussed in a series of papers \cite{Bobev:2022eus,Bobev:2022wem,Bobev:2023lkx,Bobev:2024mqw,Geukens:2024zmt,Hong:2024} for different types of partition functions such as the squashed 3-sphere partition function, TTI, and SCI. Notably, many of these partition functions are interconnected through the Bethe formulation. Therefore, we expect that gaining an analytic understanding of the Bethe vacuum beyond the strict large $N$ limit will be a crucial step toward the analytic derivation of the all order closed-form expressions. 

An important assumption underlying the Bethe formula (\ref{Z}) is that all Bethe vacua are properly isolated, allowing the partition function to be expressed as a discrete sum over them \cite{Closset:2017zgf}. It is therefore crucial to validate this assumption to confirm the Bethe formulation, or one may need to refine the formula to incorporate the contributions from continuous families of solutions to the Bethe equations. In a similar vein, the Bethe-Ansatz formula for the TTI/SCI of 4d $\mN=4$ SU($N$) Super-Yang-Mills theory \cite{Benini:2015noa,Hong:2018viz,Benini:2018mlo,Benini:2018ywd} is indeed contaminated by the presence of continuous families of solutions to the Bethe-Ansatz-Equations (BAE) \cite{ArabiArdehali:2019orz,Lezcano:2021qbj,Benini:2021ano}, and in this case the BAEs also allow for multiple isolated solutions \cite{Hong:2018viz}. Given this result in 4d, it would be essential to analyze the Bethe equations for various 3d supersymmetric theories of interest including ABJM theory, investigating the existence of additional isolated Bethe vacua that may differ from the established ones \cite{Benini:2015eyy,Liu:2017vll,Bobev:2022eus,Hosseini:2016ume,Bobev:2023lkx} as well as any continuous families of Bethe vacua. Understanding their gravity duals through the AdS/CFT correspondence would then be a natural follow-up question.

Various supersymmetric partition functions of 3d $\mN\geq 2$ holographic SCFTs with U(1)$_R$ symmetry arising from the $N$ stack of M2-branes can be expressed as a sum of higher-derivative gravitational blocks over fixed points in a dual asymptotically Euclidean AdS$_4$ background \cite{Hristov:2021qsw,Hristov:2022lcw,Hristov:2024cgj,Genolini:2021urf}. For ABJM theory, in particular, the gravitational block is proposed explicitly in terms of an Airy function \cite{Hristov:2022lcw}. This all order factorization formula aligns perfectly with the closed-form expressions for the TTI/SCI of ABJM theory derived via numerical analysis \cite{Bobev:2022jte,Bobev:2022eus,Bobev:2022wem}. However, for the $\mathcal{M}_{\mg,p}$ partition function of ABJM theory with a holographic dual Euclidean AdS-Taub-Bolt background examined in this paper, the same all order factorization formula does not apply --- except in the special case of $p=0$ corresponding to the TTI \cite{Hristov:2021qsw,Hristov:2022lcw} --- due to the presence of a two-dimensional fixed submanifold, referred to as a bolt, which is distinct from a fixed point. On the other hand, recent developments in equivariant localization \cite{BenettiGenolini:2023kxp,BenettiGenolini:2023ndb,Martelli:2023oqk,Colombo:2023fhu} led to the localization formula for the gravitational free energy around the Euclidean AdS-Taub-Bolt background in the semi-classical limit \cite{BenettiGenolini:2024xeo}. Notably, our findings suggest that the $\mathcal{M}_{\mg,p}$ partition function of ABJM theory, whose leading order in the large $N$ limit is reproduced by the aforementioned localization formula through holography \cite{BenettiGenolini:2024xeo,BenettiGenolini:2024hyd}, allows for a compact closed-form expression to all orders in the $1/N$-perturbative expansion. Thus, it would be very interesting to explore whether the localization formula for the gravitational free energy around the Euclidean AdS-Taub-Bolt background can be refined to address subleading corrections in the $1/N$-perturbative expansion, or in a similar context the higher-derivative gravitational block formula concerning ABJM theory can be improved to accommodate extended fixed submanifolds within a given background.

In this manuscript, we restrict our attention to the $1/N$-perturbative expansion of the $\mM_{\mg,p}$ partition function of ABJM theory, neglecting non-perturbative corrections of order $\mO(e^{-\sqrt{N}})$. For a global Euclidean AdS$_4$ background, such non-perturbative corrections have been analyzed from both the field theory perspective \cite{Drukker:2010nc,Hatsuda:2012dt,Nosaka:2015iiw,Nosaka:2024gle} and the string/M theory perspective via 1-loop calculations around the instanton configuration \cite{Gautason:2023igo,Beccaria:2023ujc}, with successful comparisons made between the two approaches. Exploring the same question in the context of holographic duality between the $\mM_{\mg,p}$ partition function and the dual Euclidean AdS-Taub-Bolt backgrounds would therefore be a natural following question.

In section \ref{sec:holo}, we conducted a holographic comparison focusing on the universal twist that is dual to the Euclidean AdS-Taub-Bolt backgrounds in the 4d $\mN=2$ minimal gauged supergravity. Given that the all order $\mM_{\mg,p}$ partition function of ABJM theory has been derived for generic values of holonomies and magnetic fluxes associated with flavor symmetries, it is natural to extend this comparison beyond the universal/minimal case. This extension has been explored at the $N^{3/2}$ leading order in \cite{BenettiGenolini:2024xeo,BenettiGenolini:2024hyd} through equivariant localization \cite{BenettiGenolini:2023kxp,BenettiGenolini:2023ndb,Martelli:2023oqk,Colombo:2023fhu}, without explicitly constructing a new Euclidean AdS-Taub-Bolt solution in the 4d $\mN=2$ gauged STU model. Constructing such a background characterized by non-trivial configurations of vector fields associated with the three vector multiplets, however, would certainly be an interesting project by itself, putting the holographic matching on a more solid ground. For similar constructions concerning the $S^3$ partition function and the $S^1\times\Sigma_{\mg}$ TTI on the dual field theory side, see \cite{Freedman:2013oja,Bobev:2020pjk} respectively.

\section*{Acknowledgments}

We are grateful to Nikolay Bobev, Pietro Benetti Genolini, Kiril Hristov, and Valentin Reys for valuable discussions and carefully reviewing the draft. JH is supported by the National Research Foundation of Korea (NRF) grant funded by the Korea government (MSIT) with grant number RS-2024-00449284, the Sogang University Research Grant of 202410008.01, the Basic Science Research Program of the National Research Foundation of Korea (NRF) funded by the Ministry of Education through the Center for Quantum Spacetime (CQUeST) with grant number NRF-2020R1A6A1A03047877.

\appendix

\section{Conventions}\label{app:conventions}
\noindent\textbf{3d conventions}

\medskip

A 3d spinor transforming as a doublet of SU(2), denoted by $\psi_\alpha$, has an SU(2) subscript $\alpha\in\{1,2\}$. The spinor indices can be raised by the left multiplication of the anti-symmetric rank-2 tensor as $\psi^\alpha=\varepsilon^{\alpha\beta}\psi_\beta$ with the convention $\varepsilon^{12}=1$. All spinor products are then understood in the Northwest-Southeast convention as 
\begin{equation}
	\psi\chi=\psi^\alpha\chi_\alpha\,.
\end{equation}
We choose the 3d gamma matrices $\gamma^a$ as (the Roman indices $a,b,c$ represent the local orthonormal frame indices)
\begin{equation}
	\gamma^a=\sigma^a\,,
\end{equation}
where $\sigma^a~(a\in\{1,2,3\})$ are the Pauli matrices. The components of the gamma matrices read $(\gamma)_\alpha{}^\beta$ in terms of SU(2) indices. With this convention, it becomes straightforward to derive the relation (\ref{chi:tchi}) as
\begin{equation}
	K=\partial_\psi\quad\to\quad K^a=\zeta \gamma^a\tzeta=\varepsilon^{\alpha\beta}\zeta_\beta(\sigma^a)_\alpha{}^\gamma\tzeta_\gamma=(0,0,\chi\tchi)=e^a{}_\psi=(0,0,\beta_\text{3d})
\end{equation}
for the 3d Killing spinors (\ref{KS}).

\medskip

\noindent\textbf{4d conventions}

\medskip

For the 4d gamma matrices in Euclidean signature, we employ
\begin{equation}
	\gamma^a=\Bigg\{\begin{pmatrix}
		0 & \sigma^1 \\
		\sigma^1 & 0
	\end{pmatrix}\,,~\begin{pmatrix}
	0 & \sigma^2 \\
	\sigma^2 & 0
	\end{pmatrix}\,,~\begin{pmatrix}
	0 & \sigma^3 \\
	\sigma^3 & 0
	\end{pmatrix}\,,~ \begin{pmatrix}
		0 & \ri I_2 \\
		-\ri I_2 & 0
	\end{pmatrix}\Bigg\}
\end{equation}
where the Roman indices again denote the local orthonormal frame indices in the 4d background. The Killing spinor equation (\ref{KSE}) presented in \cite{Romans:1991nq} is originally formulated in Lorentzian signature, but it retains the same structure in terms of the above Euclidean gamma matrices after performing a Wick rotation.

\bibliography{Mgp-AllOrder}
\bibliographystyle{JHEP}


\end{document}